\documentclass[%
12pt,
reprint,
onecolumn,
 amsmath,amssymb,amsthm,
 aps,
]{revtex4-2}

\usepackage[final]{graphicx}
\usepackage{setspace}
\usepackage{exscale}
\usepackage{natbib}
\usepackage{textcomp}

\usepackage{isomath}
\usepackage{amsmath}
\usepackage{amsbsy}
\usepackage{amssymb}
\usepackage{amscd}
\usepackage{amsfonts}
\usepackage{stmaryrd}
\usepackage{units}
\usepackage{euscript}
\usepackage[T1]{fontenc}
\usepackage{times}
\everymath{\displaystyle}
\usepackage{exscale}

\usepackage{graphicx}
\usepackage{boxedminipage}
\usepackage{calc}
\usepackage[usenames,dvipsnames]{xcolor}
\graphicspath{ {media/} }

\usepackage[small,compact]{titlesec}
\usepackage{enumitem}
\setitemize{noitemsep,topsep=0pt,parsep=0pt,partopsep=0pt}
\setenumerate{noitemsep,topsep=0pt,parsep=0pt,partopsep=0pt}
\setdescription{noitemsep,topsep=0pt,parsep=0pt,partopsep=0pt}

\usepackage[colorinlistoftodos, color=green!40,prependcaption]{todonotes}

\numberwithin{equation}{section}
\usepackage[,colorlinks,urlcolor=blue,citecolor=blue]{hyperref}
\usepackage[normalem]{ulem}
\usepackage[plain]{fancyref}

\usepackage{setspace}
\usepackage{isomath}
\usepackage{amsmath}
\usepackage{amssymb}
\usepackage{amscd}
\usepackage{amsfonts}

\newcommand{\half}{\frac{1}{2}}

\newcommand{\bfsigma}{\boldsymbol {\sigma}}

\newcommand{\bfalpha}{\boldsymbol {\alpha}}

\newcommand{\bftau}{\boldsymbol {\tau}}

\newcommand{\trace}{\mathop{\rm tr}\nolimits}

\newcommand{\bfe}{{\boldsymbol e}}

\newcommand{\beq}{\begin{equation}}
\newcommand{\eeq}{\end{equation}}

\newcommand{\bfB}{{\boldsymbol B}}
\newcommand{\bfC}{{\boldsymbol C}}

\newcommand{\bfE}{{\boldsymbol E}}
\newcommand{\bls}{\boldsymbol}
\newcommand{\bfF}{{\boldsymbol F}}

\newcommand{\bfI}{{\boldsymbol I}}

\newcommand{\bfQ}{{\boldsymbol Q}}

\newcommand{\bfV}{{\boldsymbol V}}

\newcommand{\bfX}{{\boldsymbol X}}

\begin{document}
	
\baselineskip=20pt

\newcommand{\titlescript}{Responses of any arbitrary initially stressed reference and the stress-free reference}
\title{\titlescript}
\author{
	Soumya Mukherjee\footnote{Email: \url{soumyamechanics@gmail.com}, \url{s.mukherjee@iitpkd.ac.in} \\ \textcolor{red}{Please see the published version}} \\Department of Mechanical Engineering, \\ Indian Institute of Technology Palakkad\\ Kerala- 678623, India}
\date{\today}
\begin{abstract}
The constitutive relation for an initially stressed reference is often determined by using the response of a virtual stress-free reference. However, identifying the constitutive relation of the original stress-free body can be challenging without conducting destructive tests. This paper presents three approaches for determining the response of a stress-free reference---or any arbitrary initially stressed reference---when the response of a particular initially stressed reference is known. Unlike standard practice, these approaches of changing reference configurations do not begin with a known stress-free state. The first and third approaches directly derive the constitutive relations of one stressed reference from another. {The first approach is applicable to a specific constitutive relation of the known initially stressed state,} while the third approach extends the first and is applicable to any constitutive form. The second approach uses any general response of a given stressed reference to identify the stress-free material. The response of the stress-free material is further analyzed and processed to determine the response of any stressed reference. We observe that even when the known initially stressed state is Green elastic, the arbitrarily stressed or stress-free references may exhibit implicit elasticity. Several explicit constitutive relations are also exemplified. {One of the examples utilizes the invariants of Seth's generalized strain measures. For experimental validation of this model, we transform the Treloar data to various initially stressed references. Model parameters corresponding to a specific stressed state are optimized by corroborating the model against the Treloar data associated with that reference. These optimized parameters are then used to determine the constitutive relations for the stress-free and other stressed reference configurations. The resulting constitutive relations for various references show identical alignment with the corresponding Treloar data, thereby validating the present approaches for changing the reference configuration.} It is demonstrated that the developed models satisfy all restrictions related to the change of references. Furthermore, we develop universal relations for stress-free isotropic implicit elastic materials.
\end{abstract}

\maketitle
\section{Introduction}
Materials are almost always subject to \emph{initial stress} in their reference
state. The mechanics of initially stressed
materials~\cite{johnson1993dependence,johnson1995use,saravanan2008representation,merodio2013influence,gower2015initial,agosti2018constitutive,mukherjee2024representation,mukherjee2024representing}
has emerged as a significant area of research. The present study develops new
frameworks for modeling the response of arbitrarily stressed (as well as
stress-free) reference configurations from the known response of a specific
initially stressed reference. \textcolor{black}{The developed frameworks
are validated against the well-known Treloar data.} Interestingly, it is
observed that changing the reference configuration often turn a Green elastic
material into an implicit elastic
material~\cite{rajagopal2003implicit,rajagopal2007elasticity,rajagopal2011spherical,rajagopal2012extension,bustamante2020review,devendiran2017thermodynamically}.
This paper establishes universal
relations~\cite{rajagopal1987new,beatty1987class,rajagopal2014universal} for
implicit elasticity. 

Hoger~\cite{hoger1985residual} determined the residual stress fields that are
admissible in elastic bodies possessing material symmetries. Johnson and
Hoger~\cite{johnson1995use,johnson1993dependence} were the first to derive the
mechanical response of an elastic material from an initially stressed
reference, wherein the \emph{virtual} stress-free configuration is modeled as
an incompressible isotropic Mooney--Rivlin solid. To achieve this, they
inverted the constitutive relation of stress-free Mooney--Rivlin materials,
obtaining the initial strain as a function of the initial stress.
Saravanan~\cite{saravanan2008representation} established the representation of
stress from an initially stressed reference, where the response of the
\emph{virtual} stress-free material is governed by the compressible isotropic
Blatz--Ko model. 

Shams, Merodio, and
co-workers~\cite{shams2011initial,merodio2013influence,merodio2016extension,shariff2017spectral,jha2019computational,jha2019constitutive,rodriguez2016helical}
also constructed the free energy of an elastic body from a stressed reference
using the invariants~\cite{shams2011initial} of the initial stress and the
right Cauchy--Green stretch tensor, treating the initial stress as a symmetry
tensor.  Many of the above models were applied to solve boundary value
problems~\cite{merodio2013influence,merodio2016extension,mukherjee2021extended},
stability analysis~\cite{melnikov2021bifurcation,desena2021computational},
\textcolor{black}{for introducing
growth~\cite{mukherjee2025deformation}} and investigating wave
propagation~\cite{shams2011initial,nam2016effect} through an initially stressed
medium. 

Destrade and
Coworkers~\cite{gower2015initial,agosti2018constitutive,mukherjee2022representing}
used a \emph{virtual} stress-free configuration to represent the strain energy
using initially stressed references. Their model has been employed to solve
numerous physical problems involving initial residual
stresses~\cite{du2018modified,mukherjee2021static,du2019influence,du2019prescribing,ciarletta2016morphology,ciarletta2016residual}.
\textcolor{black}{Du
\textit{et al.}~\cite{du2018modified,du2019influence,du2019prescribing} introduced growth
into this framework, which effectively captures the coupling between growth and
initial stress.}
Mukherjee~\cite{mukherjee2022constitutive,mukherjee2022influence} inverted the
constitutive relations of the Gent model and the Volokh
model~\cite{volokh2007hyperelasticity} to determine the response of an
initially stressed, stretch-limited material and developed a failure model for
residually stressed materials. The inverse problem of calculating the initial
strain from the initial stress was explored for nonlinear anisotropic
materials~\cite{mukherjee2022representing,mukherjee2023some}, where this
inverted constitutive relation was used to express stress and strain energy
from a stressed reference. Mukherjee and
Ravindran~\cite{mukherjee2024representation} represented stress and free energy
for viscoelastic materials using an initially stressed state. Implicit elastic
constitutive relations have been developed from an initially stressed reference
utilizing the response of a virtual stress-free
state~\cite{mukherjee2024representing}.

In the present modeling, implicit
elasticity~\cite{rajagopal2003implicit,rajagopal2007response,rajagopal2009class,rajagopal2011spherical}
often appears inevitably, while changing the reference configuration in the most
general case. Implicit
elasticity~\cite{rajagopal2003implicit,gokulnath2017representations,rajagopal2011spherical}
represents a more general material response, characterized by an implicit
constitutive relation between stress and strain. This paper develops a
universal relation for implicit elasticity, which is useful for the current
analysis. In general, universal
relations~\cite{rajagopal1987new,beatty1987class} reveal intriguing features of
constitutive relations in an elegant manner. Rajagopal and
Wineman~\cite{rajagopal2014universal} developed the universal relation for a
specific class of implicit elastic solids. In this work, we establish a
universal relationship for another general class of implicit elasticity, which
aids in the present investigation.

To model initially stressed materials, the response of the virtual stress-free body is typically required. However, identifying the stress-free response of the material can be a challenging task that often requires destructive testing. \textcolor{black}{Since materials are frequently found in an initially stressed state, it is more practical to determine their constitutive relations experimentally, based on an available stressed reference configuration. Using the experimentally determined constitutive relation of the available stressed reference, this paper derives the constitutive relations for all other reference configurations.}

\textcolor{black}{We propose the following alternative approach for modeling all
initially stressed (and the stress-free) references, which is
validated and processed using experimental data. First, the constitutive
relation for a known initially stressed state $\mathfrak{R}_1$ is determined through
corroboration with experimental observations. This relation is then used to
determine the material response for an arbitrarily stressed reference
$\mathfrak{R}$, as well as for the stress-free reference configuration $\mathfrak{R}_0$,
using new approaches for changing reference configurations.} Identifying the
constitutive relation of the stress-free reference $\mathfrak{R}_0$ from that of a stressed
reference configuration $\mathfrak{R}_1$ is converse of the conventional
approach~\cite{johnson1995use,saravanan2008representation,mukherjee2023some},
in which the response of a stressed reference $\mathfrak{R}$ is derived
using a virtual stress-free reference $\mathfrak{R}_0$.

We present a few approaches for changing reference configurations to determine the response of any arbitrary reference $\mathfrak{R}$ from the response of the reference $\mathfrak{R}_1$. New types of constitutive relations are inverted. The first approach directly and easily determines the response of the reference $\mathfrak{R}$ for a particular constitutive relation of the reference $\mathfrak{R}_1$. The second approach uses any general response of $\mathfrak{R}_1$ to sequentially determine the response of the stress-free reference $\mathfrak{R}_0$ and the arbitrary stressed reference $\mathfrak{R}$. Quite interestingly, we observe that the general constitutive relations of the references $\mathfrak{R}_0$ and $\mathfrak{R}$ may demonstrate implicit elasticity, while $\mathfrak{R}_1$ exhibits Green elasticity. Some special examples of Green elastic constitutive relations are identified for references $\mathfrak{R}_0$ and $\mathfrak{R}$. The third approach generalizes the first approach and determines ways to develop more general constitutive relations for the reference $\mathfrak{R}$.

\textcolor{black}{Furthermore, a generalized free energy function, involving fractional powers of the right Cauchy stretch, is considered for reference $\mathfrak{R}_1$. Using the present framework, the stress and strain energies are derived using all other stressed references, $\mathfrak{R}$, and the stress-free reference, $\mathfrak{R}_0$. This constitutive relation is used for experimental validation.}

\textcolor{black}{For experimental validation, we transform the famous
Treloar data to various initially stressed references. A two-parameter model is
used to corroborate the experimental results for the reference $\mathfrak{R}_1$. By
processing the optimized parameters for reference $\mathfrak{R}_1$, the responses of
all other references are uncovered. The resulting constitutive relations for
all references---$\mathfrak{R}$, $\mathfrak{R}_1$, and $\mathfrak{R}_0$---show
\emph{identical} agreement with the corresponding experimental results. This
observation ensures that our current approach accurately alters the reference
configurations. It is noted that optimizing only two parameters yields
satisfactory agreement with a large set of experimental data for several
initially stressed (and stress-free) reference configurations. By slightly
increasing the number of parameters in the present framework----for example, by
adopting a form similar to the 3-parameter Carroll
model~\cite{carroll2011strain,steinmann2012hyperelastic} for reference
$\mathfrak{R}_1$, the present frameworks can achieve excellent agreements with the
huge set of Treloar data for uniaxial and biaxial tension, as well as for pure
shear, simultaneously for all initially stressed and stress-free references.}

Although it is not the main focus of this paper, we briefly examine whether the functional forms of the developed constitutive relations depend on the choice of reference configuration. We demonstrate that our frameworks are consistent with all relevant theories in this regard. 

We also establish universal relations for general implicit elasticity. 

This paper is organized as follows. In  Section \ref{given}, we present the constitutive relation used for the given stressed reference $\mathfrak{R}_1$ (with an initial stress $\bftau_1$). We then proceed to transform the reference configuration, selecting any arbitrary initially stressed state or a stress-free state as the reference. 

 Section \ref{2nd} describes the first approach, which directly determines the response of an arbitrary stressed reference $\mathfrak{R}$ (with initial stress $\bftau$) using the given stressed reference $\mathfrak{R}_1$ (with initial stress $\bftau_1$). This approach is easily applied for a particular constitutive relation for $\mathfrak{R}_1$.

   Our second approach (Section \ref{1st}) uses any general known response of the stressed reference $\mathfrak{R}_1$ to sequentially determine the response of the stress-free reference $\mathfrak{R}_0$ (Section \ref{stress--free}) and any arbitrary initially stressed reference $\mathfrak{R}$ (Section \ref{resres}). In the various subsections, we establish (a) general implicit responses for both references, $\mathfrak{R}$ and $\mathfrak{R}_0$, and (b) two special explicit constitutive relations for $\mathfrak{R}$ and $\mathfrak{R}_0$.

The third approach (Section \ref{3rd}) generalizes the first approach by introducing an \emph{imaginary} initially stressed reference $\mathfrak{R}_1'$, constructed by rotating the initial stress field of $\mathfrak{R}_1$ in a certain manner. It should be noted that $\mathfrak{R}_1'$ is not a real configuration. This imaginary configuration enables an easy inversion of the constitutive relation of the known stressed state. This method is particularly useful when the reference $\mathfrak{R}_1$ is governed by a more general or intricate constitutive relation. 

\textcolor{black}{ Section \ref{gen_model} presents the transformation of the reference configuration when the strain energy density relative to the stressed reference configuration, $\mathfrak{R}_1$, is expressed in terms of the invariants of the fractional powers of the right Cauchy stretch. The developed model is validated with experimental data in  Section \ref{Comapre}.} 

\textcolor{black}{For experimental validation, in  Section \ref{Data_Traloar}, we obtain the Treloar data corresponding to various initially stressed references. For a particular reference configuration $\mathfrak{R}_1$, the data is corroborated using a two parameter model (Section \ref{corroboration}). The resulting optimized pair of parameters are used to derive the models all other references $\mathfrak{R}$ and $\mathfrak{R}_0$ which exhibit consistent agreement with the corresponding Treloar data.}

In  Section \ref{ogd-art}, we study whether, according to the present framework, the functional form of the constitutive relation depends on the choice of reference or not. We find that both dependency and independence are satisfied.

\textcolor{black}{In  Section \ref{uni}, universal relations are developed for a particular class of implicit elasticity.}
\section{The known response of an initially stressed reference $\mathfrak{R}_1$}
\label{given}
Continuum mechanics cherishes the assumption of a stress-free reference. However, in real-life situations, a stress-free configuration is usually unattainable. The stress in the reference configuration is known as the initial stress \cite{hoger1985residual}. 
A general approach to model an initially stressed reference can be obtained in \cite{shams2011initial,merodio2013influence,merodio2016extension}.

In this section, using the existing invariants of initial stress and the right Cauchy stretch, we present the constitutive relation for a given initially stressed reference, $\mathfrak{R}_1$, with the initial stress denoted as $\bftau_1$. The material parameters associated with the strain energy density are determined by corroborating experimental data (Section \ref{corroboration}).

When the initial stress $\bftau_1(\bfX_1)$ does not involve any traction on the boundary of $\mathfrak{R}_1$, it is known as a residual stress field.

\begin{figure}
    \centering
    \includegraphics[scale=0.98]{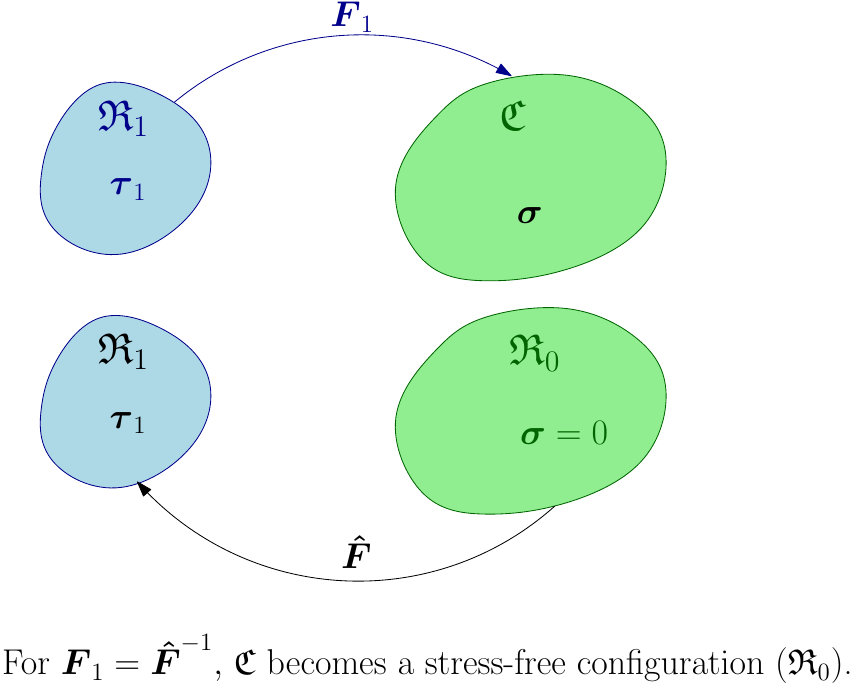}
    \caption{\doublespacing The configurations required in Section \ref{given} and Section \ref{stress--free}. \textbf{Top:} the deformation gradient $\bfF_1$ maps a stressed reference $\mathfrak{R}_1$ to the current configuration $\mathfrak{C}$, with Cauchy stress $\boldsymbol{\sigma}$.
    \textbf{Bottom:} the deformation gradient $\hat{\bfF}$ maps a vector from the \emph{stress-free reference} $\mathfrak{R}_0$ to the configuration $\mathfrak{R}_1$. This deformation $\hat{\bfF}$ is associated with initial/ residual strain. \emph{Note:} when Cauchy stress $\bfsigma=0$ in the current configuration, $\mathfrak{C}$ is equivalent to $\mathfrak{R}_0$. This particular case presents us access with the stress-free reference without destructive testing.}
    \label{fig1}
\end{figure}
In Figure~\ref{fig1} (top), the reference configuration $\mathfrak{R}_1$ undergoes deformation with a gradient $\bfF_1$ to produce the current configuration $\mathfrak{C}$. The associated Helmholtz potential $\Psi$ can be expressed in terms of the following invariants:
\begin{align}
   \hspace{-1cm}& I_1= \trace\bfC_1,
   &&I_2=\trace\left(\bfC_1^2\right),
   && I_3=\text{det}\left(\bfC_1\right),
   && I_{4}=\trace\bftau_1,
   && I_{5}= \tfrac{1}{2}[I_4^2 - \trace(\bftau_1^2)], \nonumber
   \\
    \hspace{-1cm}& I_{6}=\text{det}\left(\bftau_1\right),
   && I_7= \trace\left(\bls{\tau}_1\bls{C}_1\right),
   && I_8= \trace\left(\bls{\tau}_1\bls{C}_1^2\right),
   &&  I_{9} = \trace\left( \bls{\tau}_1^2\bls{C}_1\right),
   &&  I_{10} = \trace\left( \bls{\tau}_1^2\bls{C}_1^2\right),
   \label{inv4}
\end{align}
where $\bfC_1=\bfF_1^T\bfF_1$ is the right Cauchy-Green stretch tensor. 

The Cauchy stress in the current $\mathfrak{C}$ is determined from the potential $\Psi$ as
\begin{align}
    \bfsigma& = \Psi_1 \bfB_1 + \Psi_2 \bfB_1^2 +\Psi_7 \bfF_1\bftau_1\bfF_1^T+\Psi_8\left(\bfF_1\bftau_1\bfF_1^T\bfB_1+\bfB_1\bfF_1\bftau_1\bfF_1^T\right)\nonumber\\&+\Psi_9 \bfF_1\bftau_1^2\bfF_1^T+\Psi_{10}\left(\bfF_1\bftau_1^2\bfF_1^T\bfB_1+\bfB_1\bfF_1\bftau_1^2\bfF_1^T\right)-p\bfI\label{consti}
\end{align}
where $\bfB_1=\bfF_1\bfF_1^T$, $\Psi_i=\textcolor{black}{2}\tfrac{\partial \Psi}{\partial I_i}$, and $p$ is the Lagrangian multiplier associated with the incompressibility constraint. 

Since the initial stress is defined as the stress in the reference configuration, substituting$\bfF_1=\bfI$ into Eqn. \eqref{consti} yields:
\begin{align}
    &\bftau_1=\left(\Psi_{1_0}+\Psi_{2_0}-p_0\right)\bfI+\left(\Psi_{7_0}+2\Psi_{8_0}\right)\bftau_1+\left(\Psi_{9_0}+2\Psi_{{10}_0}\right)\bftau_1^2,\label{eee}
\end{align}
where $\Psi_{i_0}=\Psi_{i}\Big|_{\bfF=\bfI}$, $p_{0}=p\Big|_{\bfF=\bfI}$

Eqn. \eqref{consti} represents the constitutive relation for the stressed reference $\mathfrak{R}_1$. For a given initial stress $\bftau_1$, this analytical form \eqref{consti} can be used to corroborate experimental results, as demonstrated in Section \ref{corroboration}.

Some of our goals in this paper are to identify (a) the response of the stress-free reference $\mathfrak{R}_0$, (b) the initial strain stored in the given reference $\mathfrak{R}_1$, (c) the representation of Cauchy stress for any arbitrarily stressed reference $\mathfrak{R}$ with initial stress $\bftau$, and (d) to provide an experimental validation of the developed frameworks.

All of these goals are accomplished by using the known response of a single stressed reference $\mathfrak{R}_1$ through different approaches in Section \ref{2nd}, Section \ref{1st}, and Section \ref{3rd}, respectively.
Section \ref{1st} and Section \ref{gen_model} present examples of constitutive relations for various references. In Section \ref{Comapre}, the current frameworks are validated using experimental data.

\section{\textcolor{black}{A simplified approach to identifying the response of the stress-free reference $\mathfrak{R}_0$ and any stressed reference $\mathfrak{R}$}}
\label{2nd}
\begin{figure}[hbt!]
    \centering
    \includegraphics[scale=.85]{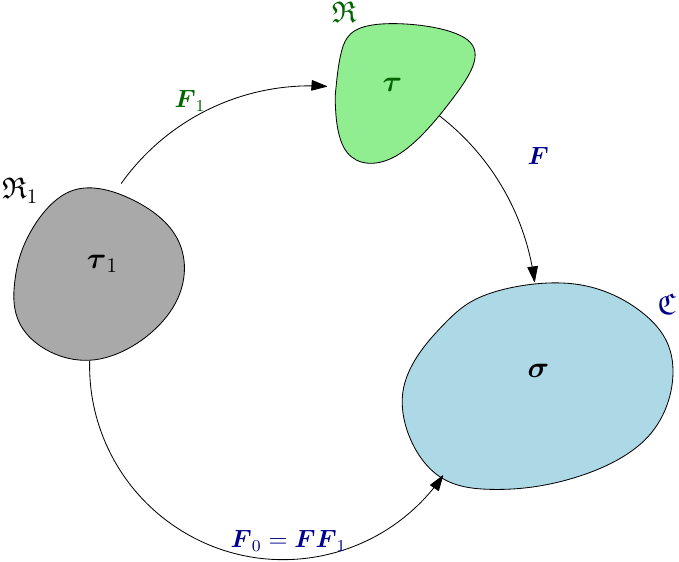}
    \caption{\doublespacing
    The constitutive relation for the initially stressed configurations $\mathfrak{R}_1$, with initial stress $\bftau_1$, is known. The goal is to find the response from the reference $\mathfrak{R}$ (where the initial stress $\bftau$ can be arbitrary).}
    \label{fig:my_label}
\end{figure}

\textcolor{black}{In this section, we examine the response of a special constitutive relation relative to the reference configuration $\mathfrak{R}_1$} to directly determine the constitutive response for an arbitrarily stressed reference $\mathfrak{R}$ with an arbitrary initial stress $\bftau$ (see Figure \ref{fig:my_label}). 
The three configurations $\mathfrak{R}_1$, $\mathfrak{R}$, and $\mathfrak{C}$ with Cauchy stress $\bftau_1$, $\bftau$, and $\bfsigma$, respectively, are illustrated in
Figure \ref{fig:my_label}. The deformation gradients $\bfF_1$, $\bfF$, and $\bfF_0$ transform any infinitesimal vector from $\mathfrak{R}_1\to\mathfrak{R}$, $\mathfrak{R}\to\mathfrak{C}$, and $\mathfrak{R}_1\to\mathfrak{C}$, respectively, such that $\bfF_0=\bfF\bfF_1$. The reference $\mathfrak{R}_1$ follows the constitutive relation \eqref{consti}, for which
\begin{equation}
{\Psi}_1=p_0, \quad {\Psi}_2={\Psi}_8={\Psi}_9={\Psi}_{10}=0, \quad {\Psi}_7=1. \label{cit112}
\end{equation}
When $\mathfrak{R}_1$ is chosen as a reference, the Cauchy stress $\bfsigma$ depends on $\bfF_1$ and $\bftau_1$. For this choice of reference, the Cauchy stresses in the configurations $\mathfrak{C}$ and $\mathfrak{R}$ (refer to Figure \ref{fig:my_label}) are expressed as 
\begin{align} \bfsigma&=\Psi_1\bfB_0+\bfF_0\bftau_1\bfF_0^T-p\bfI,\label{actual}\\ \bftau &=\Psi_1\bfB_1+\bfF_1\bftau_1\bfF_1^T-{p}_1\bfI,\label{acty} \end{align} where $p$ and ${p}_1$ denote the Lagrangian multipliers for $\mathfrak{C}$ and $\mathfrak{R}$ respectively.

The multiplicative decomposition $\bfF_0=\bfF\bfF_1$, and subsequent simplifications reformulate Equation \eqref{actual} as follows: \begin{align} \bfsigma&=\Psi_1\bfF\bfB_1\bfF^T+\bfF\bfF_1\bftau_1\bfF_1^T\bfF^T-p\bfI\label{hjhj}\\ &=\bfF\left[\Psi_1\bfB_1+\bfF_1\bftau_1\bfF_1^T\right]\bfF^T-p\bfI.\label{actual-1} \end{align} We rewrite Equation \eqref{acty} as \begin{align} &\Psi_1\bfB_1+\bfF_1\bftau_1\bfF_1^T=\bftau+ p_1\bfI.\label{56} \end{align} Substitution of \eqref{56} into \eqref{actual-1} yields \begin{align} \bfsigma=p_1\bfB+\bfF\bftau\bfF^T-p\bfI.\label{actual---111} \end{align} The constitutive relation \eqref{actual---111} solely depends on the initial stress $\bftau$ in $\mathfrak{R}$ and the gradient $\bfF$ of deformation measured from the configuration $\mathfrak{R}\rightarrow\mathfrak{C}$. Hence, \eqref{actual---111} establishes the constitutive relation for the stressed reference $\mathfrak{R}$ that we aimed to determine.

The unknown parameter $p_1$ is derived from the initial stress $\bftau$ by calculating the determinants of both sides of Eqn. \eqref{56} as \begin{align} \text{det}\left(\bfF_1\right)\text{det}\left(\Psi_1\bfI+\bftau_1\right)\text{det}\left(\bfF_1^T\right)=\text{det}\left(\bftau+ p_1\bfI\right).\label{444} \end{align} Substitution of the incompressibility condition $\text{det}\left(\bfF_1\right)=1$ simplifies the expression \eqref{444} as, \begin{align} &\Psi_1^3+\Psi_1^2I_{4}+\Psi_1I_{5}+I_{6}=p_1^3+p_1^2I_{\tau_1}+p_1I_{\tau_2}+I_{\tau_3}\label{4545} \end{align} where $I_{4}$, $I_{5}$, and $I_{6}$ are the three invariants \eqref{inv4} of $\bftau_1$ and $I_{\tau_1}$, $I_{\tau_2}$, and $I_{\tau_3}$ denote the three invariants of $\bftau$. The parameter $p_1$ is determined by solving equation \eqref{4545}. For $\bftau=\bfQ\bftau_1\bfQ^T$, where $\bfQ$ represents a pure rotation, Equation \eqref{4545} results in \beq p_1=\Psi_1.\label{git}\eeq In this scenario of simple rotation, the constitutive relation \eqref{actual} for reference $\mathfrak{R}_1$ is identical to that of $\mathfrak{R}$.

The present derivation is applicable to any arbitrary value of the initial stress, $\bftau$. For $\bftau=\bf0$, the solution to Equation \eqref{4545} is given by
\begin{equation}
\bar{p}=\sqrt[3]{\Psi_1^3+\Psi_1^2I_{4}+\Psi_1I_{5}+I_{6}},
\end{equation}
and \eqref{actual---111} reduces to
\begin{align}
\bfsigma=\bar{p}\bfB-p\bfI.\label{actual-111}
\end{align}
This denotes the constitutive response of the corresponding stress-free material.

This section outlines a simple and direct approach to identify a \emph{specific} constitutive relation for the references $\mathfrak{R}$ and $\mathfrak{R}_0$. This approach is extended and generalized in Section \ref{3rd}. Section \ref{1st} presents our most general framework of changing reference configuration, where the constitutive relation may become implicit due to changes in the reference configuration.
\section{The second approach of changing the reference configuration}
\label{1st}
This section presents a generalized approach wherein the reference $\mathfrak{R}_1$ is allowed to follow any constitutive relation of the form given in Eq.~\eqref{consti}, and determined the constitutive relations for both the stress-free reference $\mathfrak{R}_0$ and the arbitrary stressed reference configuration $\mathfrak{R}$, (with initial stress $\bftau$).

The general constitutive response of the stress-free reference $\mathfrak{R}_0$ is derived in Section \ref{stress--free}, including a range of special cases.
Section \ref{resres} determines the constitutive relations.
Section \ref{resres} investigates the constitutive relations for an arbitrarily stressed reference $\mathfrak{R}$ through various approaches. In both cases, the general response may be implicit, as we observe in this section.



\subsection{Identifying the response of the stress-free reference}
\label{stress--free}
This section evaluates the response of stress-free material by inverting an interesting constitutive relation. Appendix A establishes universal relations \cite{rajagopal1987new,beatty1987class,rajagopal2014universal} for a new implicit constitutive class, which aids in simplifying the analysis.
\subsubsection{The general implicit constitutive relation for stress-free materials}
\label{sss}
Obtaining a stress-free material state without conducting a destructive testing is challenging. In this work, we employ a thought experiment, as depicted in Figure \ref{fig1}, to access the stress-free configuration.

Figure \ref{fig1} (top) shows that a deformation gradient $\bfF_1$ maps the stressed configuration $\mathfrak{R}_1$ to the current configuration $\mathfrak{C}$, where the Cauchy stress is given by $\bfsigma$. When $\bfsigma$ vanishes, the current configuration $\mathfrak{C}$ is equivalent to the stress-free configuration, $\mathfrak{R}_0$ (Figure \ref{fig1} (bottom)). In this scenario, we have the relation
\begin{equation}
  \bfF_1 = \hat{\bfF}^{-1}, \label{ff}
\end{equation}
where $\hat{\bfF}$ (Figure \ref{fig1} (bottom)) represents the gradient of deformation from the stress-free configuration $\mathfrak{R}_0$ to the stressed reference $\mathfrak{R}_1$. The deformation gradient $\hat{\bfF}$ is associated with the initial strain in $\mathfrak{R}_1$. Equation \eqref{ff} further leads to the following relations:
\begin{align}
\bfC_1 = \hat{\bfB}^{-1}, \quad \bfB_1 = \hat{\bfC}^{-1}, \label{bb}
\end{align}
where $\hat{\bfB} = \hat{\bfF} \hat{\bfF}^T$, and $\hat{\bfC} = \hat{\bfF}^T \hat{\bfF}$ are alternative measures of residual strain in the reference $\mathfrak{R}_1$, while $\bfB_1 = \bfF_1 \bfF_1^T$ and $\bfC_1 = \bfF_1^T \bfF_1$. It is important to note that the stress-free configuration $\mathfrak{R}_0$ can undergo any arbitrary rotation while maintaining its stress-free state. Consequently, the stress-free configuration is not unique. However, this non-uniqueness does not impact the present analysis, as justified in Appendix \ref{apex}.
 
To conduct the thought experiment outlined above, we select $\mathfrak{R}_1$ as the reference configuration, governed by the constitutive relation \eqref{consti}, and $\mathfrak{C}$ as the current configuration. In this context, we have the condition
\begin{equation}
    \bfsigma = 0. \label{0sigma}
\end{equation}
Substituting equations \eqref{ff}, \eqref{bb}, and \eqref{0sigma} into \eqref{consti}, we obtain the following expression:
\begin{align}
    \bf0 & = \bar{\Psi}_1 \hat{\bfC}^{-1} + \bar{\Psi}_2 \hat{\bfC}^{-2} + \bar{\Psi}_7 \hat{\bfF}^{-1} \bftau_1 \hat{\bfF}^{-T} + \bar{\Psi}_8 \left( \hat{\bfF}^{-1} \bftau_1 \hat{\bfF}^{-T} \hat{\bfC}^{-1} + \hat{\bfC}^{-1} \hat{\bfF}^{-1} \bftau_1 \hat{\bfF}^{-T} \right) \nonumber \\
    & \quad + \bar{\Psi}_9 \hat{\bfF}^{-1} \bftau_1^2 \hat{\bfF}^{-T} + \bar{\Psi}_{10} \left( \hat{\bfF}^{-1} \bftau_1^2 \hat{\bfF}^{-T} \hat{\bfC}^{-1} + \hat{\bfC}^{-1} \hat{\bfF}^{-1} \bftau_1^2 \hat{\bfF}^{-T} \right) - \bar{p} \bfI, \label{consti-6}
\end{align}
where $\bar{\Psi}_i = \Psi_i \big|_{\bfF_1 = \hat{\bfF}^{-1}}$ and $\bar{p} = p \big|_{\bfF_1 = \hat{\bfF}^{-1}}$.

Next, by pre-multiplying equation \eqref{consti-6} by $\hat{\bfF}$ and post-multiplying by $\hat{\bfF}^T$, we simplify the equation to
\begin{align}
    \bf0& = \bar{\Psi}_1 \bfI + \bar{\Psi}_2 \hat{\bfB}^{-1} +\bar{\Psi}_7 \bftau_1+\bar{\Psi}_8\left(\bftau_1\hat{\bfB}^{-1}+\hat{\bfB}^{-1}\bftau_1\right)+\bar{\Psi}_9 \bftau_1^2+\bar{\Psi}_{10}\left(\bftau_1^2\hat{\bfB}^{-1}+\hat{\bfB}^{-1}\bftau_1^2\right)-\bar{p}\hat{\bfB}.\label{consti-7}
\end{align}

As shown in Figure \ref{fig1} (bottom), we now choose $\mathfrak{R}_0$ as the \emph{stress-free reference} and $\mathfrak{R}_1$ as the \emph{current} configuration. In this case, $\hat{\bfF}$, $\hat{\bfB}$, and $\bftau_1$ denote the deformation gradient, the left Cauchy-Green stretch tensor, and the Cauchy stress in the current configuration $\mathfrak{R}_1$, respectively.

It is interesting to note that Equation \eqref{consti-7} represents the constitutive relation for the stress-free reference $\mathfrak{R}_0$. It correlates the Cauchy stress $\left(\bftau_1\right)$ with the left Cauchy stretch $\left(\hat{\bfB}\right)$. This general constitutive relation \eqref{consti-7} is implicit in nature and is equivalent to and closely resembles the standard form of the implicit constitutive relation \cite{rajagopal2007elasticity,rajagopal2003implicit,bustamante2020review,gokulnath2017representations}.

The constitutive relation \eqref{consti-7} clearly illustrates that the stress-free reference $\mathfrak{R}_0$ exhibits isotropy. The universal relations for these general implicit elastic materials are derived in Appendix \ref{uni}, which can further simplify \eqref{consti-7} in certain cases. It should be noted that Rajagopal and
Wineman~\cite{rajagopal2014universal} established universal relations for another class of implicit elasticity. From the universal relations presented in Appendix \ref{uni}, we have
\begin{equation}
  \hat{\bfB}^{-1}\bftau_1=  \bftau_1\hat{\bfB}^{-1}, \quad  \hat{\bfB}^{-1}\bftau_1^2=\bftau_1^2\hat{\bfB}^{-1}.\label{lll}
\end{equation}
which can express \eqref{consti-7} as
\begin{align}
    \bf0& = \left(\bar{\Psi}_1 \bfI+\bar{\Psi}_7 \bftau_1 +\bar{\Psi}_9 \bftau_1^2\right)\bfI -\bar{p}\hat{\bfB}+ \left(\bar{\Psi}_2 \bfI +2\bar{\Psi}_8\bftau_1+2\bar{\Psi}_{10}\bftau_1^2\right)\hat{\bfB}^{-1}, \text{ or}\label{consti-8}\\
 \bf0& =\left(\bar{\Psi}_1 \bfI -\bar{p}\hat{\bfB}+\bar{\Psi}_2\hat{\bfB}^{-1}\right){\bfI}+ \left(\bar{\Psi}_7 \bfI +2\bar{\Psi}_8\hat{\bfB}^{-1}\right){\bftau_1} + \left(2\bar{\Psi}_{10} \bfI +\bar{\Psi}_9\hat{\bfB}\right){\bftau_1^2} \label{consti-11}\
\end{align}

It is noted that when $\bar{\Psi}_i\left(\bfX_1\right)$ are independent of $\bftau_1$ or $\hat{\bfF}_1$, we have \beq
{\Psi}_i\left(\bfX_1\right)=\bar{\Psi}_i\left(\bfX_1\right)={\Psi}_{i_0}\left(\bfX_1\right).\label{eqality}
\eeq (Also see Eqn. \ref{eee}.)

Equations (\ref{consti-7}, \ref{consti-8}, \ref{consti-11}) represent the general implicit relation between the Cauchy stress $\bftau_1$ and the left Cauchy stretch $\hat{\bfB}$ for the stress-free reference $\mathfrak{R}_0$. Two special explicit forms of this constitutive relation for $\mathfrak{R}_0$ are derived in Section \ref{next} and Section \ref{next-1}, respectively.
\subsubsection{Special case-I}
\label{next}
For the special case 
\begin{equation}
\bar{\Psi}_1=p_0, \quad \bar{\Psi}_2=\bar{\Psi}_8=\bar{\Psi}_9=\bar{\Psi}_{10}=0. \quad \bar{\Psi}_7=1, \label{cit}
\end{equation}
the implicit constitutive relation \eqref{consti-8} for the stress-free reference $\mathfrak{R}_0$ reduces to the following explicit form:
\begin{equation}
    \bftau_1= \bar{p}\hat{\bfB}-\bar{\psi}_1\bfI=\bar{p}\hat{\bfB}-p_0\bfI.\label{cit2}
\end{equation}
The parameter $\bar{\Psi}_1=p_0$ serves as the Lagrange multiplier for incompressibility during this change of reference.
Inverting \eqref{cit2}, we determine the initial strain $\hat{\bfB}$ in $\mathfrak{R}_1$ as
\begin{equation}
   \hat{\bfB}=\frac{1}{\bar{p}}\bftau_1+\frac{p_0}{\bar{p}}\bfI.\label{fgf}
\end{equation}
The parameter $\bar{p}$ is determined by setting the determinant of both the sides of \eqref{fgf} equal to unity (due to incompressibility) to obtain
\begin{equation}
    \bar{p}=\sqrt[3]{\text{det}\left[\bftau_1+\bar{\Psi}_1\bfI\right]}.\label{jh}
\end{equation}


The constitutive relation \eqref{cit2} is applicable when $\mathfrak{R}_0$ is chosen as reference and $\mathfrak{R}_1$ as the current. Extending \eqref{cit2}, the general constitutive relation for the \emph{stress-free reference} $\mathfrak{R}_0$ and a current configuration $\mathfrak{C}_1$ (see Figure \ref{fi22}) is obtained as
\begin{equation}
    \bfsigma=\bar{p}{\bfB}_{tot}-p\bfI.\label{coo}
\end{equation}
where $\bfB_{tot}=\bfF_{tot}\bfF_{tot}^T$, $\bfF_{tot}$ maps any vector from $\mathfrak{R}_0$ to a current configuration $\mathfrak{C}_1$. At this stage a new deformation gradient $\bfF_{tot}$ and a new current configuration $\mathfrak{C}_1$ (Figure \ref{fi22}) are introduced. Once the response of the stress-free reference is found, the reference $\mathfrak{R}_1$ is no longer required to further change the reference. 
We use the constitutive relation \eqref{coo} for $\mathfrak{R}_0$ to identify the response of an arbitrary initial stressed reference $\mathfrak{R}$ (in Section \ref{nextc}). 
\subsubsection{Special case-II}
\label{next-1}
Here we consider the special case
\begin{equation}
\bar{\Psi}_2=p_0, \quad \bar{\Psi}_1=\bar{\Psi}_7=\bar{\Psi}_9=\bar{\Psi}_{10}=0, \quad \bar{\Psi}_8=\half, \label{citt}
\end{equation}
for which the constitutive relation \eqref{consti-7}
\begin{align}
    \bf0& = \bar{\Psi}_2 \hat{\bfB}^{-1}+\bar{\Psi}_8\left(\bftau_1\hat{\bfB}^{-1}+\hat{\bfB}^{-1}\bftau_1\right)-\bar{p}\hat{\bfB}.\label{consti-77}
\end{align}
is simplified into
\begin{align}
    \bftau_1& =\frac{\bar{p}}{2\bar{\Psi}_8}\hat{\bfB}^2 -\frac{\bar{\Psi}_2}{2\bar{\Psi}_8}\bfI,\label{consti-788}
\end{align}
through a few intermediate steps and {the application of universal relations (Appendix \ref{uni})}. 
In the constitutive relation \eqref{consti-788}, $\tfrac{\bar{\Psi}_2}{2\bar{\Psi}_8}$ functions as the Lagrange multiplier, with $\mathfrak{R}_0$ representing the reference state and $\mathfrak{R}_1$ referring to the current state. The initial strain in $\mathfrak{R}_1$ is evaluated from \eqref{consti-788} as
\begin{align}
\hat{\bfB}=\sqrt{\left(\frac{2\bar{\Psi}_8}{\bar{p}}\bftau_1+\frac{\bar{\Psi}_2}{\bar{p}}\bfI\right)},
\end{align}
and $\bar{p}$ can be determined as
\begin{equation}
    \bar{p}=\sqrt[3]{\text{det}\left[2\bar{\Psi}_8\bftau_1+\bar{\Psi}_2\bfI\right]}.\label{jh-1}
\end{equation}
using the incompressibility condition $\text{det}\hat{\bfB}=1$.

By extending the constitutive relation \eqref{consti-788} for the \emph{stress-free reference} $\mathfrak{R}_0$, the Cauchy stress $\bfsigma$ in the current $\mathfrak{C}_1$ (illustrated in Figure \ref{fi22}), is expressed as follows,
\begin{equation}
    \bfsigma=\frac{\bar{p}}{2\bar{\Psi}_8}{\bfB}^2_{tot}-p\bfI,\label{chekeat}
\end{equation}
where $\bfB_{tot}=\bfF_{tot}\bfF_{tot}^T$ (refer to Figure \ref{fi22}). 
We use the relations developed in this section to determine the response of the initially stressed reference $\mathfrak{R}$ in Section \ref{nextc-1}
\subsection{\label{resres}The response of any arbitrarily stressed reference $\mathfrak{R}$}
In this section, our goal is to determine the response of any arbitrarily stressed reference $\mathfrak{R}$ with initial stress $\bftau$. Consequently, constitutive relations and methodologies established in Sections \ref{sss}, \ref{next}, and \ref{next-1} are used and further extended in Sections \ref{checksec}, \ref{nextc}, and \ref{nextc-1}, respectively. 
\subsubsection{An implicit constitutive relation for the stressed reference $\mathfrak{R}$}
\label{checksec}
\begin{figure}
    \centering
    \includegraphics[scale=1.3]{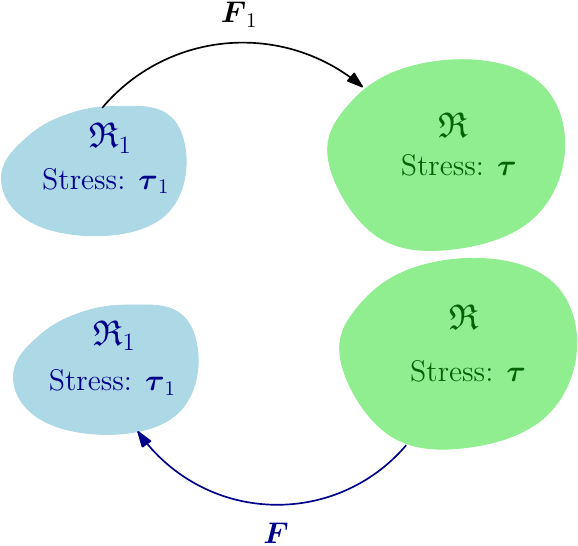}
    \caption{\doublespacing The strategy for addressing the solution in Section \ref{checksec}. \textbf{Top:} we select $\mathfrak{R}_1$ as the reference, and $\mathfrak{R}$ as the current, where the deformation gradient is $\bfF_1$. \textbf{Bottom:} $\mathfrak{R}$ is chosen as the reference and $\mathfrak{R}_1$ as the current, with deformation gradient $\bfF=\bfF_1^{-1}$. The constitutive relation \eqref{consti} is known for the case at the Top and determined for the case at the Bottom.}
    \label{fig23}
\end{figure}

In contrast to the approach in Section \ref{sss}, we consider that the current configuration $\mathfrak{C}$ is not stress-free. The Cauchy stress in $\mathfrak{C}$ is represented by $\bftau$, which implies that $\mathfrak{C}\equiv\mathfrak{R}$.  The configurations are illustrated schematically in Fig. \ref{fig23} (top). Choosing \eqref{consti} as the constitutive relation for reference $\mathfrak{R}_1$, the Cauchy stress in $\mathfrak{R}$ is expressed as
\begin{align}
    \bftau& = \Psi_1 \bfB_1 + \Psi_2 \bfB_1^2 +\Psi_7 \bfF_1\bftau_1\bfF_1^T+\Psi_8\left(\bfF_1\bftau_1\bfF_1^T\bfB_1+\bfB_1\bfF_1\bftau_1\bfF_1^T\right)\nonumber\\&+\Psi_9 \bfF_1\bftau_1^2\bfF_1^T+\Psi_{10}\left(\bfF_1\bftau_1^2\bfF_1^T\bfB_1+\bfB_1\bfF_1\bftau_1^2\bfF_1^T\right)-\textcolor{black}{p_1}\bfI,\label{subst}
\end{align}
\textcolor{black}{where $p_1$ represents the Lagrange multiplier associated with the deformation from the reference configuration $\mathfrak{R}_1$ to the current configuration $\mathfrak{R}$.}
As illustrated in Figure \ref{fig23} (bottom), we will now consider $\mathfrak{R}$ as the initially stressed reference and $\mathfrak{R}_1$ as the current state, with the Cauchy stress represented by $\bftau_1$. In this section, the deformation gradient from $\mathfrak{R}$ to $\mathfrak{R}_1$ is represented by $\bfF$.
We substitute $\bfF_1=\bfF^{-1}$ and $\bfB_1=\bfC^{-1}$ into Eqn. \eqref{subst}, then pre-multiply the resulting equation by ${\bfF}$ and post-multiply it by ${\bfF}^T$, to obtain
\begin{align}
    0& = \bar{\Psi}_1 \bfI + \bar{\Psi}_2 {\bfB}^{-1} +\bar{\Psi}_7 \bftau_1+\bar{\Psi}_8\left(\bftau_1{\bfB}^{-1}+{\bfB}^{-1}\bftau_1\right)+\bar{\Psi}_9 \bftau_1^2+\bar{\Psi}_{10}\left(\bftau_1^2{\bfB}^{-1}+{\bfB}^{-1}\bftau_1^2\right)-{\bfF}\bftau{\bfF}^T-\textcolor{black}{p_1}{\bfB}.\label{consti-78}
\end{align}
Eqn. \eqref{consti-78} represents the implicit constitutive relation for the arbitrary stressed reference $\mathfrak{R}$ (with initial stress $\bftau$), with $\mathfrak{R_1}$ as the current configuration (with Cauchy stress $\bftau_1$). 

Mukherjee \cite{mukherjee2024representing} introduced an implicit elastic framework for initially stressed reference configurations. The present constitutive relation Eqn. \eqref{consti-78} belongs to the general form presented in Eqn. (2.2) of \cite{mukherjee2024representing}. However, Mukherjee \cite{mukherjee2024representing} has not derive a constitutive relation of the present form by changing the reference configuration. 

This section yields an implicit elastic elastic constitutive relation for the reference $\mathfrak{R}$. Section \ref{next,next-1} present the Green elastic response of the stress-free reference
configuration, $\mathfrak{R}_0$. Building on this, we next determine the Green elastic constitutive
relation for $\mathfrak{R}$ by extending the approach of Johnson and
Hoger~\cite{johnson1993dependence,johnson1995use},
Saravanan~\cite{saravanan2008representation,saravanan2011large}, and Gower \textit{et al.} \cite{gower2015initial}. Notably, the configuration $\mathfrak{R}_1$ is no longer required for this purpose. Instead, a new current configuration, $\mathfrak{C}_1$, is introduced in Figure \ref{fi22}. 
\subsubsection{A Green elastic constitutive relation for the stressed reference $\mathfrak{R}$}
\label{nextc}
In this section, we derive the constitutive relation for the initially stressed reference $\mathfrak{R}$, provided that the stress-free configuration $\mathfrak{R}_0$ is governed by the constitutive relation \eqref{coo} (Section \ref{next}).
\begin{figure} 
    \centering
    \includegraphics[scale=0.75]{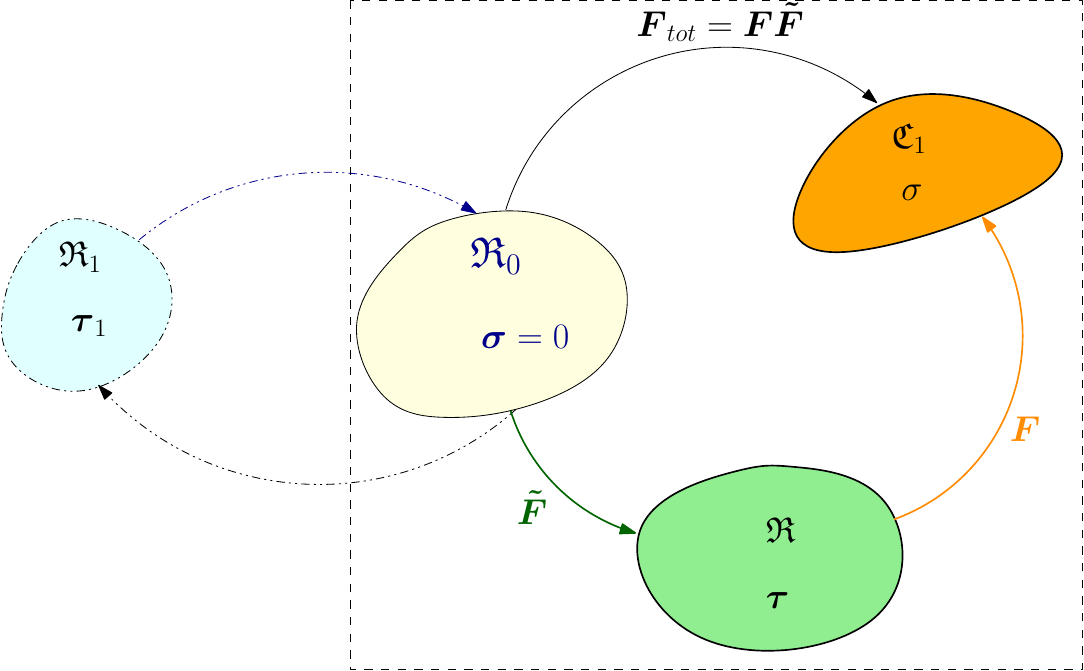}
    \caption{\doublespacing Using the response of the stress-free reference $\mathfrak{R}_0$, derived in Sections \ref{next} and \ref{next-1}, the constitutive relations for an arbitrarily stressed reference $\mathfrak{R}$ are formulated in Sections \ref{nextc} and \ref{nextc-1}. The configuration $\mathfrak{R}_1$ is no longer needed in this process. \textcolor{black}{The deformation gradient $\boldsymbol{F}_{tot}$ can be expressed as $\boldsymbol{\bfF\tilde{F}}$ through multiplicative decomposition.}}
    \label{fi22}
\end{figure}

Figure \ref{fi22} describes the stress-free configuration $\mathfrak{R}_0$, initially stressed reference $\mathfrak{R}$, and the current configuration $\mathfrak{C}_1$. The deformation gradient from $\mathfrak{R}_0\rightarrow\mathfrak{R}$ is $\tilde{\bfF}$, from $\mathfrak{R}\rightarrow\mathfrak{C}_1$ is $\bfF$, and from $\mathfrak{R}_0\rightarrow\mathfrak{C}_1$ is given by  $\bfF_{tot}=\bfF\tilde{\bfF}$. 
The Cauchy stress in configuration $\mathfrak{R}$ is denoted as $\bftau$, while in configuration $\mathfrak{C}_1$, it is represented by $\bfsigma$.

Applying the constitutive relation \eqref{coo} for the reference $\mathfrak{R}_0$,  the current configuration is chosen respectively as $\mathfrak{R}$ and $\mathfrak{C}_1$, we obtain
\begin{align}
\bftau&=\bar{p}\boldsymbol{\tilde{B}}-\textcolor{black}{p_1}\bfI, \label{coo1}\\
\bfsigma&=\bar{p}\bfF_{tot}\bfF_{tot}^T-p\bfI=\bar{p}\bfF\boldsymbol{\tilde{B}}\bfF^T-p\bfI,\label{coo2}
\end{align}
where the multiplicative decomposition $\bfF_{tot}=\bfF\tilde{\bfF}$ is used. 
Eqn. \eqref{coo1} can also be alternatively expressed as
\begin{equation}
\bar{p}\boldsymbol{\tilde{B}}=\bftau+\textcolor{black}{p_1}\bfI.\label{coo-1}
\end{equation}

\textcolor{black}{Eqn. \eqref{coo-1} is substituted into Eqn. \eqref{coo2}} to obtain
\begin{equation}
\bfsigma=\bfF\bftau\bfF^T+\textcolor{black}{p_1}\bfB-p\bfI.\label{ccc--1}
\end{equation}
The constitutive relation \eqref{ccc--1} depends on the the initial stress $\bftau$ in the reference $\mathfrak{R}$, and on the deformation gradient $\bfF$ from $\mathfrak{R}$ to $\mathfrak{C}_1$. Thus, it denotes the constitutive relation for the arbitrarily stressed reference $\mathfrak{R}$.
The unknown parameter $\textcolor{black}{p_1}$ is determined in terms of $\bftau$ as follows.
Calculating the determinant on both the sides of \eqref{coo-1} and substituting the incompressibility condition $\text{det}\tilde{\bfB}=1$, we obtain
\begin{equation}
\bar{p}^3=\text{det}\left(\bftau+\textcolor{black}{p_1}\bfI\right),\label{now}
\end{equation}
which can be solved for $\textcolor{black}{p_1}$, once $\bar{p}$ is obtained from Equation \eqref{fgf}.A relation similar to Eqn. \eqref{now} was derived in \cite{gower2015initial}. \textcolor{black}{Furthermore, note that the parameter $p_1$ in Eqns. \eqref{now} and \eqref{444} is identical.}

\subsubsection{A different Green elastic constitutive relation for the stressed reference $\mathfrak{R}$.}
\label{nextc-1}
When the stress-free reference $\mathfrak{R}_0$ abides by the constitutive relation Eqn. \eqref{chekeat} (Section \ref{next-1}), we determine the response of the reference $\mathfrak{R}$ with initial stress $\bftau$ (Figure \ref{fi22}). Substituting $\bfF_{tot}=\bfF\tilde{\bfF}$ into Eqn. \eqref{chekeat} with simplifications, we obtain
\begin{equation}
\bfsigma=\frac{\bar{p}}{2\bar{\Psi}_8}\bfF\tilde{\bfB}\bfC\tilde{\bfB}\textcolor{black}{{\bfF}^T}-p\bfI.\label{chekeat-1}
\end{equation}
On the other hand, while using $\mathfrak{R}_0$ as the reference and $\mathfrak{R}$ as the current configuration (see Figure \ref{fi22}), the deformation gradient is given by $\tilde{\boldsymbol{F}}$. The Cauchy stress in the configuration $\mathfrak{R}$ takes the form
\begin{equation}
\bftau=\frac{\bar{p}}{2\bar{\Psi}_8}{\tilde{\bfB}}^2-\textcolor{black}{p_1}\bfI,\label{chekeat-99}
\end{equation}
where $\textcolor{black}{p_1}$ is the Lagrange multiplier for the configuration $\mathfrak{R}$.

Eqn. \eqref{chekeat-99} can be inverted as
\beq
\tilde{\bfB}=\sqrt{\left(\frac{2\bar{\Psi}_8}{\bar{p}}\bftau+\frac{2\bar{\Psi}_8\textcolor{black}{p_1}}{\bar{p}}\bfI\right)}\label{invrt}
\eeq

We can substitute Eqn. \eqref{invrt} into Eqn. \eqref{chekeat-1} to express the Cauchy stress relative to an arbitrarily stressed reference \(\mathfrak{R}\). The Lagrange multiplier \(\textcolor{black}{p_1}\) is determined by enforcing the incompressibility condition \(\text{det} \tilde{\bfB} = 1\) in Eqn. \eqref{invrt}, after computing \(\bar{p}\) from Eqn. \eqref{jh-1}.  

In this section, we observe that the general constitutive relation for either the stress-free reference \(\mathfrak{R}_0\) or an initially stressed reference \(\mathfrak{R}\) can be implicit. This demonstrates the significance and generality of the implicit elasticity \cite{rajagopal2007elasticity,rajagopal2003implicit,rajagopal2010new,bustamante2020review}. For two specific cases, explicit constitutive relations are also determined.
\section{The third approach of changing the reference configuration}
\label{3rd}
The method described in Section \ref{1st} establishes a general framework to change the reference configuration from one initially stressed state to another.
The first approach (Section \ref{2nd}) is direct and straightforward. which is direct and straightforward, applies to a specific choice of constitutive relation relative to the reference $\mathfrak{R}_1$. This simple approach can serve as motivation and presents a foundation for exploring more general approaches. To extend this approach, in Section \ref{AB}, we examine some challenges associated with its applications to more intricate cases. Subsequently, in Section \ref{gen_gen}, we extend and generalize this approach to develop more general constitutive relations for the stressed reference $\mathfrak{R}$.

\subsection{Challenges in generalizing the first Approach}
\label{AB}
We applied the approach of Section \ref{2nd} when the response of the reference $\mathfrak{R}_1$ is governed by the constitutive relation Eqn. \eqref{actual}. When the constitutive relation of reference $\mathfrak{R}_1$ is expressed in a more general and intricate form as presented in Eqn. \eqref{consti}, the Cauchy stresses in the configurations $\mathfrak{C}$ and $\mathfrak{R}$ (Figure \ref{fig:my_label}) are computed as 
\begin{align}
    \bfsigma& = \Psi_1 \bfB_0 + \Psi_2 \bfB_0^2 + \Psi_7 \bfF_0 \bftau_1 \bfF_0^T + \Psi_8 \left( \bfF_0 \bftau_1 \bfF_0^T \bfB_0 + \bfB_0 \bfF_0 \bftau_1 \bfF_0^T \right)\nonumber \\
    &+ \Psi_9 \bfF_0 \bftau_1^2 \bfF_0^T + \Psi_{10} \left( \bfF_0 \bftau_1^2 \bfF_0^T \bfB_0 + \bfB_0 \bfF_0 \bftau_1^2 \bfF_0^T \right) - p \bfI, \label{aactual} \\
    \bftau& = \Psi_1 \bfB_1 + \Psi_2 \bfB_1^2 + \Psi_7 \bfF_1 \bftau_1 \bfF_1^T + \Psi_8 \left( \bfF_1 \bftau_1 \bfF_1^T \bfB_1 + \bfB_1 \bfF_1 \bftau_1 \bfF_1^T \right)\nonumber \\
    &+ \Psi_9 \bfF_1 \bftau_1^2 \bfF_1^T + \Psi_{10} \left( \bfF_1 \bftau_1^2 \bfF_1^T \bfB_1 + \bfB_1 \bfF_1 \bftau_1^2 \bfF_1^T \right) - p_1 \bfI, \label{aacty}
\end{align}
The equations \eqref{aactual} and \eqref{aacty} represent the generalized forms of Eqns. \eqref{actual} and \eqref{acty}, respectively.
By substituting the multiplicative decompositions: $\bfF_0 = \bfF \bfF_1$, $\bfB_0 = \bfF \bfB_1 \bfF^T$, and $\bfB_0^2 = \bfF \bfB_1 \bfC \bfB_1 \bfF^T$ into Eqn. \eqref{aactual}, we obtain
\begin{align}
    \bfsigma& = \Psi_1 \bfF\bfB_1\bfF^T + \Psi_2 \bfF\bfB_1\bfC\bfB_1\bfF^T +\Psi_7 \bfF\bfF_1\bftau_1\bfF_1^T\bfF^T\nonumber\\&+\Psi_8\left(\bfF\bfF_1\bftau_1\bfF_1^T\bfC\bfB_1\bfF^T+\bfF\bfB_1\bfC\bfF_1\bftau_1\bfF_1^T\bfF\right)+\Psi_9 \bfF\bfF_1\bftau_1^2\bfF_1^T\bfF^T\nonumber\\&+\Psi_{10}\left(\bfF\bfF_1\bftau_1^2\bfF_1^T\bfC\bfB_1\bfF^T+\bfF\bfB_1\bfC\bfF_1\bftau_1^2\bfF_1^T\bfF\right)-p\bfI,\label{scit}
\end{align}
 which generalizes Eqn. \eqref{hjhj}. To represent stress $\bfsigma$ from reference $\mathfrak{R}$, it is essential to express $\bfsigma$ solely as a function of $\bfF$ and $\bftau$ (see Figure \ref{fig:my_label}), thereby removing the initial strain-dependent terms $\bfB_1$, $\bfF_1\bftau_1\bfF_1^T$, and $\bfF_1\bftau_1^2\bfF_1^T$ from Eqn. \eqref{scit}. Eliminating these terms using \eqref{aacty} in the approach described in Section \ref{2nd} is challenging.
 
It is also difficult to evaluate $\bfF_1$ by inverting Eqn. \eqref{aacty} so that the initial strain-dependent terms can be eliminated from Eqn. \eqref{scit}. This difficulty arises due to the fact that $\bfF_1$ possesses $9$ independent components, whereas Eqn. \eqref{aacty} provides only $6$ simultaneous equations in three-dimensional space. The three additional components stem from the non-unique rotation part of $\bfF_1$.

To address this issue, in Section \ref{gen_gen}, we introduce a method which exclusively involves six unknown initial strain components, which can be determined from a set of six equations. This formulation enables the successful elimination of the initial strain-dependent terms, ensuring a well-posed and practically feasible solution.

\subsection{Generalizing the approach of Section \ref{2nd}}
\label{gen_gen}
To avoid the above challenges associated with the non-unique rotation part of $\bfF_1$ (Section \ref{AB}), we introduce an \emph{imaginary} stressed reference $\mathfrak{R}_1'$ which ensures that there are only $6$ unknown components of the deformation gradient $\bfF_1$, eliminating the non-unique rotation. 

\begin{figure}
    \centering
    \includegraphics[scale=0.85]{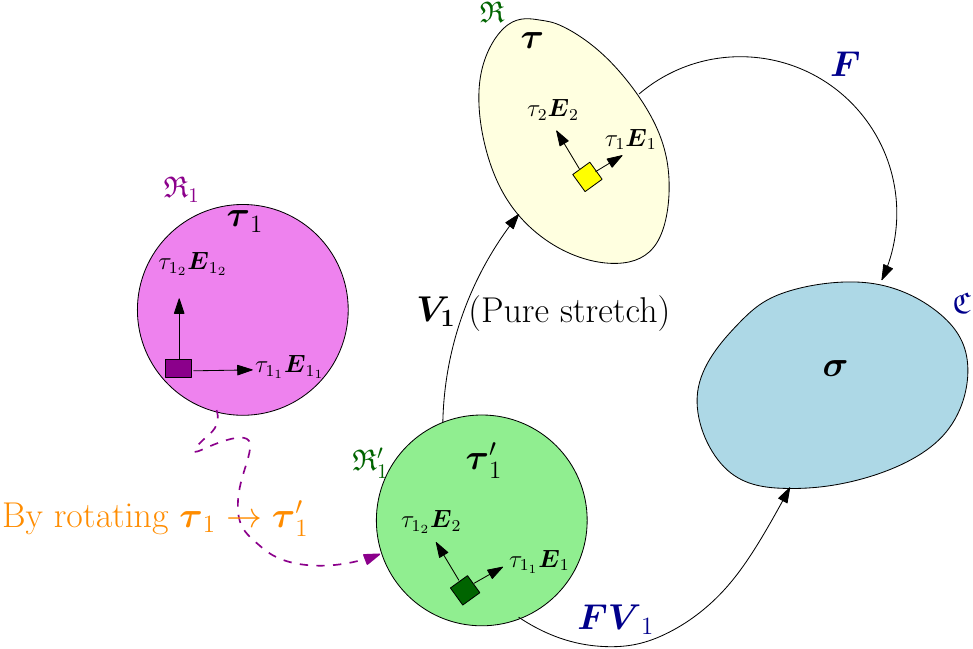}
    \caption{\doublespacing The general approach of Section \ref{3rd}. The initial stress $\bftau_1$ in $\mathfrak{R}_1$ is transformed to $\bftau_1'=\bfQ\bftau_1\bfQ^T$ resulting in an imaginary stressed configuration $\mathfrak{R}_1'$, such that the stress $\bftau_1'$ becomes coaxial with $\bftau$ in $\mathfrak{R}$. Due to this co-axiality and given that the stress-free material is isotropic, the deformation from $\mathfrak{R}_1'$ to $\mathfrak{R}$ must be a pure stretch $\bfV_1$, coaxial with both $\bftau_1'$ and $\bftau$. This pure stretch has six independent components determined by inverting the associated constitutive relation. Since $\mathfrak{R}_1'$ represents an imaginary configuration that cannot and need not be physically accessed, the stress-field $\bftau_1'$ is not required to be self-balancing. Only the kinematic relations and the constitutive relation are required in the present context.}
    \label{seeeit}
\end{figure}

This approach can be effectively visualized (see Figure \ref{seeeit}) by diagonalizing $\bftau$ and $\bftau_1$ as follows:
\begin{align}
&\bftau_1=\sum_i\tau_{1_i}{\bfE_1}_i\otimes {\bfE_1}_i &&\bftau=\sum_i\tau_{i}{\bfE}_i\otimes {\bfE}_i,
\end{align}
where ${\bfE_1}_i$, $\bfE_i$ represent the eigenvectors, and $\tau_{1_i}$, $\tau_i$ are the eigenvalues of the initial stress $\bftau_1$ and $\bftau$ respectively.

An \emph{imaginary} stressed reference $\mathfrak{R}'_1$ is constructed (Figure \ref{seeeit}) that is subject to an initial stress
\begin{equation}
\bftau_1'=\sum_i\tau_{1_i}{\bfE}_i\otimes {\bfE}_i.
\end{equation}
The principal directions of $\bftau_1'$ align with the eigenvectors of $\bftau$ and the principal values of $\bftau_1'$ are identical to the eigenvalues of $\bftau_1$. It is important to note that $\mathfrak{R}_1'$ is an \emph{imaginary} configuration, which is used only to determine the constitutive relation for $\mathfrak{R}$. The stress field $\bftau'$ need not be self-balancing since this field is \emph{not} assigned to any real reference configuration. 

Since the initial stress $\bftau_1'$ (in $\mathfrak{R}_1'$) and the Cauchy stress $\bftau$ (in $\mathfrak{R}$) are co-axial, and there is no source of anisotropy other than residual stress, the deformation from $\mathfrak{R}_1'$ to $\mathfrak{R}$ must be a \emph{pure stretch} $\bfV_1$, coaxial with both $\bftau$ and $\bftau_1'$, expressed as follows: \begin{equation} \bfV_1=\lambda_1^2{\bfE}_i\otimes {\bfE}_i. \label{vvv} \end{equation}

It should be noted that the stress \(\bftau_1'\) can be derived from the following orthogonal transformation of \(\bftau_1\):

\[
\bftau_1' = \bfQ \bftau_1 \bfQ^T \quad \text{where} \quad \bfQ = \sum_i \mathbf{E}_i \otimes \mathbf{E}_{1_i}.
\]

Since the stress-free reference is isotropic, a mere rotation of the initial stress does \emph{not} alter the response of the stressed reference (see Eqn. \eqref{git} and the subsequent discussions). Consequently, the response of the imaginary reference \(\mathfrak{R}_1'\) will also be governed by the constitutive relation presented in Eqn. \eqref{consti} (equivalently, Eqns. \eqref{aacty} and \eqref{scit}) for the reference \(\mathfrak{R}_1\).
%

By selecting \(\mathfrak{R}_1'\) as the reference, we substitute \(\bftau_1\) with \(\bftau_1'\) and \(\bfF_1\) with \(\bfV_1\) in Eqns. \eqref{aacty} and \eqref{scit}. We then replace \(\bfV_1^2\) with \(\bfB_1\) and utilize the coaxiality of \(\bfV_1\) and \(\bftau_1'\) to compute the Cauchy stresses in the configurations \(\mathfrak{R}\) and \(\mathfrak{C}\) as follows:
\begin{align}
    \bftau& = -p_1\bfI+\bfB_1\left(\Psi_1\bfI+\Psi_7\bftau_1'+\Psi_9\bftau_1'^2\right) + \bfB_1^2 \left(\Psi_2\bfI+2\Psi_8\bftau_1'+2\Psi_{10}\bftau_1'^2\right).\label{a_acty}\\
    \bfsigma& = \Psi_1 \bfF\bfB_1\bfF^T + \Psi_2 \bfF\bfB_1\bfC\bfB_1\bfF^T +\Psi_7 \bfF\bftau_1'\bfB_1\bfF^T+\Psi_8\bfF\left(\bftau_1'\bfB_1\bfC\bfB_1+\bfB_1\bfC\bfB_1\bftau_1'\right)\bfF^T\nonumber\\&+\Psi_9 \bfF\bftau_1'^2\bfB_1\bfF^T+\Psi_{10}\bfF\left(\bftau_1'^2\bfB_1\bfC\bfB_1+\bfB_1\bfC\bfB_1\bftau_1'^2\right)\bfF^T-p\bfI,\label{scitt}
\end{align}
Note here that $\bfB_1$ is the only unknown tensor in \eqref{a_acty} and \eqref{scitt}, which must be eliminated.

Note that \(\bfB_1\) is the only unknown tensor in Eqns. \eqref{a_acty} and \eqref{scitt}, which must be eliminated.

In three dimensions, \(\bfB_1\) has 6 independent components, and Eqn. \eqref{a_acty} presents a set of \emph{exactly} 6 equations. Therefore, \(\bfB_1\) can be fully determined by inverting Eqn. \eqref{a_acty} (following the approach of \cite{johnson1995use, saravanan2008representation, truesdell1952mechanical}, or using an alternative method). These strain components should then be substituted into Eqn. \eqref{scitt} to eliminate terms that depend on the initial strain. An example is provided as follows.
For ${\Psi}_1=p_0, \quad {\Psi}_2={\Psi}_8={\Psi}_9={\Psi}_{10}=0, \quad {\Psi}_7=1,$ Eqns. \eqref{a_acty} and \eqref{scitt} assume the form 
\begin{align}
    \bftau& = -p_1\bfI+\bfB_1\left(p_0\bfI+\bftau_1'\right),\label{lkj}\\
        \bfsigma& = \bfF\left(p_0\bfI+\bftau_1'\right)\bfB_1\bfF^T-p\bfI.\label{lkj1}
\end{align}
We can invert Eqn. \eqref{lkj} to determine
\beq\bfB_1=\left(p_0\bfI+\bftau_1'\right)^{-1}\left(p_1\bfI+\bftau\right),\label{eqql}\eeq which is substituted into Eqn. \eqref{lkj1} to obtain the constitutive relation \beq
\bfsigma = \bfF\left(p_1\bfI+\bftau\right)\bfF^T-p\bfI\label{gggg}\eeq
for the stressed reference $\mathfrak{R}$. The present approach is suitable for determining the constitutive relation of \(\mathfrak{R}\) by considering more general responses for the reference \(\mathfrak{R}_1\). To this end, we need to invert Eqn. \eqref{a_acty} by introducing an expression for \(\bftau^2\) of the following form:
\[
\bftau^2 = \bfalpha_0 \bfI + \bfalpha_1 \bfB + \bfalpha_2 \bfB^2,
\]
utilizing the Cayley-Hamilton theorem. The methodology presented in \cite{truesdell1952mechanical, saravanan2008representation, johnson1993dependence} can be extended to the present context. Alternatively, each component of $\bfB_1$ can be obtained directly by numerically solving \ref{a_acty}.
\textcolor{black}{\section{Example of a strain energy function using generalized invariants}
\label{gen_model}
This section demonstrates an application of the approach developed in Section \ref{1st}, where the Helmholtz free energy function \( \Psi \), relative to the reference configuration \( \mathfrak{R}_1 \), is expressed in terms of the following generalized invariants \cite{mukherjee2020generalized}:
\begin{align}
I_{11_i} &= \trace{\bfC_1^{p_i/2}} = \lambda_{1_1}^{p_i} + \lambda_{1_2}^{p_i} + \lambda_{1_3}^{p_i}, \nonumber \\
I_{12_i} &= \bftau_1 : \bfC_1^{r_i/2} = \left[\bftau_{1}\right]_{11} \lambda_{1_1}^{r_i} + \left[\bftau_{1}\right]_{22} \lambda_{1_2}^{r_i} + \left[\bftau_{1}\right]_{33} \lambda_{1_3}^{r_i}, \nonumber \\
I_{13_i} &= \bftau_1^2 : \bfC_1^{s_i/2} = \left[\bftau_{1}^2\right]_{11} \lambda_{1_1}^{s_i} + \left[\bftau^2_{1}\right]_{22} \lambda_{1_2}^{s_i} + \left[\bftau_{1}^2\right]_{33} \lambda_{1_3}^{s_i}, \label{inv-i}
\end{align}
where $\left[\bftau_1\right]_{ij} = \bfe_{1_i} \cdot \bftau_1 \bfe_{1_j}$, $\left[\bftau_1^2\right]_{ij} = \bfe_{1_i} \cdot \bftau_1^2 \bfe_{1_j}$ for $i,j \in \left\{1,2,3\right\}$ (see \cite{shariff2017spectral}). Here, $\lambda^2_{1_i}$ and $\bfe_{1_i}$ represent the principal values and directions of the right Cauchy-Green deformation tensor $\bfC_1$, respectively. It should be noted that these invariants can be derived from the generalized strain measures proposed by Seth \citep{united1964generalized,seth1978generalized,10.1007/978-3-662-29364-5_51,seth1962transition,Seth1978}. 
}

\textcolor{black}{
We further consider that the free energy function, relative to the reference configuration $\mathfrak{R}_1$, is given by:
\begin{align}
    \Psi= \frac{\mu_1}{q_1}\trace{\bfC_1^{q_1/2}}+\frac{1}{q_1}\bftau_1:{\bfC_1^{q_1/2}},\label{vu-1}
\end{align}
which, according to the theory developed in \cite{mukherjee2020generalized}, satisfies the initial condition:
\begin{align}
   \bfsigma\big|_{\bfF_1=\bfI}= \bftau_1,
\end{align}
where the Cauchy stress $\bfsigma$ is defined as:
\begin{align}
   \bfsigma=2\bfF_1\tfrac{\partial\psi}{\partial\bfC_1}\bfF_1^T-p\bfI,\label{si}
\end{align}
whose components in the principal Eulerian coordinates are given by:
\begin{align}
    \sigma_{11}&=\mu_1\lambda_{1_1}^{q_1}+\left[\bftau_{1}\right]_{11}\lambda_{1_1}^{q_1}-p,\nonumber\\
    \sigma_{22}&=\mu_1\lambda_{1_2}^{q_1}+\left[\bftau_{1}\right]_{22}\lambda_{1_2}^{q_1}-p,\nonumber\\
    \sigma_{33}&=\mu_1\lambda_{1_3}^{q_1}+\left[\bftau_{1}\right]_{33}\lambda_{1_3}^{q_1}-p,\nonumber\\
\sigma_{12}&=\left(\frac{2\left[\bftau_{1}\right]_{12}\lambda_{1_1}\lambda_{1_2}}{q_1}\right)\frac{\lambda_{1_1}^{q_1}-\lambda_{1_2}^{q_1}}{\lambda_{1_1}^{2}-\lambda_{1_2}^{2}},\nonumber\\
\sigma_{23}&=\left(\frac{2\left[\bftau_{1}\right]_{23}\lambda_{1_2}\lambda_{1_3}}{q_1}\right)\frac{\lambda_{1_2}^{q_1}-\lambda_{1_3}^{q_1}}{\lambda_{1_2}^{2}-\lambda_{1_3}^{2}},\nonumber\\
\sigma_{13}&=\left(\frac{2\left[\bftau_{1}\right]_{13}\lambda_{1_1}\lambda_{1_3}}{q_1}\right)\frac{\lambda_{1_1}^{q_1}-\lambda_{1_3}^{q_1}}{\lambda_{1_1}^{2}-\lambda_{1_3}^{2}}.\label{ssii}
\end{align}
We now employ this constitutive relation for the reference configuration $\mathfrak{R}_1$ to determine the response in the stress-free state.
}

\textcolor{black}{
\subsection{The constitutive relation associated with the stress-free reference}
\label{stressfree_example}
In this section, the approach of Section \ref{stress--free} is used to determine the response relative to the stress-free reference configuration $\mathfrak{R}_0$. To ensure that the current configuration is stress-free, we impose \( \bfsigma = \bf0 \) and set \( \bfF_1 = \hat{\bfF}^{-1} \) in Eqn. \eqref{si}, where \( \hat{\bfF} \) is the gradient of the initial deformation in \( \mathfrak{R}_1 \). Equivalently, we substitute \( \lambda_{1_i} = \hat{\lambda}_i^{-1} \), \( p = \bar{p} \), and \( \sigma_{ij} = 0 \) into Eqn. \eqref{ssii}, leading to the following:
\begin{align}
\left[\bftau_{1}\right]_{11}&=\bar{p}\hat{\lambda}_1^{q_1}-\mu_1,\nonumber\\    \left[\bftau_{1}\right]_{22}&=\bar{p}\hat{\lambda}_2^{q_1}-\mu_1,\nonumber\\
\left[\bftau_{1}\right]_{33}&=\bar{p}\hat{\lambda}_3^{q_1}-\mu_1,\nonumber\\
\left[\bftau_{1}\right]_{12}&= \left[\bftau_{1}\right]_{13}= \left[\bftau_{1}\right]_{23}=0.\label{ssiii}
\end{align}
The constitutive relation defined in Eqn. \eqref{ssiii} represents the material response with $\mathfrak{R}_0$ as the stress-free reference and $\mathfrak{R}_1$ as the current configuration, where $\bftau_1$ denotes the Cauchy stress. Notably, $\hat{\lambda}_i$ presents the principal stretch from $\mathfrak{R}_0$ to $\mathfrak{R}_1$. The shear stress components, $\left[\bftau_{1}\right]_{12}$, $\left[\bftau_{1}\right]_{13}$, and $\left[\bftau_{1}\right]_{23}$ vanish since the stress-free reference is isotropic, and the principal directions of left Cauchy stretch tensor are used here as the coordinate system. The isotropy of $\mathfrak{R}_0$ ensures that $\hat{\bfB}$ and $\bftau_1$ share the same set of principal directions. Consequently, the shear stress components vanish in the chosen coordinate system.}

\textcolor{black}{Interestingly, the constitutive relation in Eqn. \eqref{ssiii} is associated with the Ogden model, where $\mu_1$ acts as the Lagrange multiplier enforcing incompressibility. The shear modulus $\bar{p}$ of the stress-free reference $\mathfrak{R}_0$ is determined from the following incompressibility constraint for the reference $\mathfrak{R}_1$:
\begin{equation}
\hat{\lambda}_1^{q_1}\hat{\lambda}_2^{q_1}\hat{\lambda}_3^{q_1}=\frac{\left(\left[\bftau_{1}\right]_{11}+\mu_1\right)\left(\left[\bftau_{1}\right]_{22}+\mu_1\right)\left(\left[\bftau_{1}\right]_{33}+\mu_1\right)}{\bar{p}^3}=1.\label{vu}
\end{equation} 
The Ogden model described in Eqn. \eqref{ssiii} for reference $\mathfrak{R}_0$ can be used to determine Cauchy stress and strain energy at any current configuration $\mathfrak{C}_1$ (refer to Figure 4 of the manuscript), as outlined below:
\begin{align}
&\bfsigma=\bar{p}\bfB_{tot}^{q/2}-p\bfI, && \Psi= \frac{\bar{p}}{q_1}\left(\trace{\bfC_{tot}^{q/2}}-3\right).\label{fen}
\end{align}
An analogous expression for Cauchy stress in $\mathfrak{C}_1$ is provided in Eqn. \eqref{coo}.}
\textcolor{black}{\subsection{Response from an arbitrary reference $\mathfrak{R}$}
Figure \ref{fi22} illustrates that the arbitrary reference $\mathfrak{R}$ enables the decomposition of the deformation gradient $\bfF_{tot}$ into the product $\bfF\tilde{\bfF}$. Implementing this multiplicative decomposition, the free energy in the current configuration $\mathfrak{C}_1$ (Eqn. (\ref{fen}b)) is rewritten as,
\begin{align}
\Psi=\frac{\bar{p}}{q_1}\trace{\left(\tilde{\bfF}^T\bfC\tilde{\bfF}\right)^{q_1/2}}-3\frac{\bar{p}}{q_1}=\frac{\bar{p}}{q_1}\trace{\left(\bfC\tilde{\bfB}\right)^{q_1/2}}-3\frac{\bar{p}}{q_1},\label{ppsi}
\end{align} which explicitly involves the unknown initial strain $\tilde{\bfB}$ of the reference $\mathfrak{R}$.}

\textcolor{black}{
The initial stress $\bftau$ in the reference $\mathfrak{R}$ can be derived from the free energy function (Eqn. (\ref{fen}a)) as
\begin{align}
\bftau=\bar{p}\tilde{\bfB}^{q_1/2}-\textcolor{black}{p_1}\bfI.\label{tt}
\end{align}
By solving for $\tilde{\bfB}$ in Eqn. \eqref{tt} and substituting it into Eqn. \eqref{ppsi}, the strain energy function takes the form:
\begin{equation}
\Psi=\frac{1}{q_1}\trace\left[\bfC\left(\bftau+\textcolor{black}{p_1}\bfI\right)^{2/q_1}\right]^{q_1/2}-3\frac{\bar{p}}{q_1}.\label{developed}
\end{equation}
The parameter $\textcolor{black}{p_1}$ in Eqn. \eqref{developed} can be calculated from Eqn. \eqref{tt} through the incompressibility condition. Applying Eqn. \eqref{vu}, the material parameters corresponding to different reference configurations are related by:
\begin{equation}
\bar{p}=\sqrt[3]{\text{det}\left(\bftau_1+\mu\bfI\right)}=\sqrt[3]{\text{det}\left(\bftau+\textcolor{black}{p_1}\bfI\right)}\label{immm}
\end{equation}}
\textcolor{black}{We have developed a constitutive relation, Eqn. \eqref{developed}, for the reference $\mathfrak{R}$. This formulation provides a generalized framework for modeling the response of materials with an arbitrary initial stress state.}

\textcolor{black}{To validate the proposed model, Section \ref{Comapre} presents a detailed comparison with the well-established Treloar experimental data. This comparison demonstrates the effectiveness of the constitutive relation developed in accurately representing the mechanical behavior of any arbitrary reference configuration. 
\section{Comparison with Treloar data from various initially stressed states}
\label{Comapre}
This section optimizes the material parameters for a specific initially stressed reference $\mathfrak{R}_1$ by calibrating the analytical model \eqref{vu-1} against experimental data. These optimized parameters are then used to determine the material response of all other references, including both $\mathfrak{R}$ and $\mathfrak{R}_{0}$. It is noted that all constitutive relations derived for these different reference configurations exhibit a consistent and identical agreement with the corresponding experimental results, confirming the robustness and accuracy of the proposed formulation.}

\begin{figure}[hbt!]
    \centering
    \includegraphics[width=0.48\linewidth]{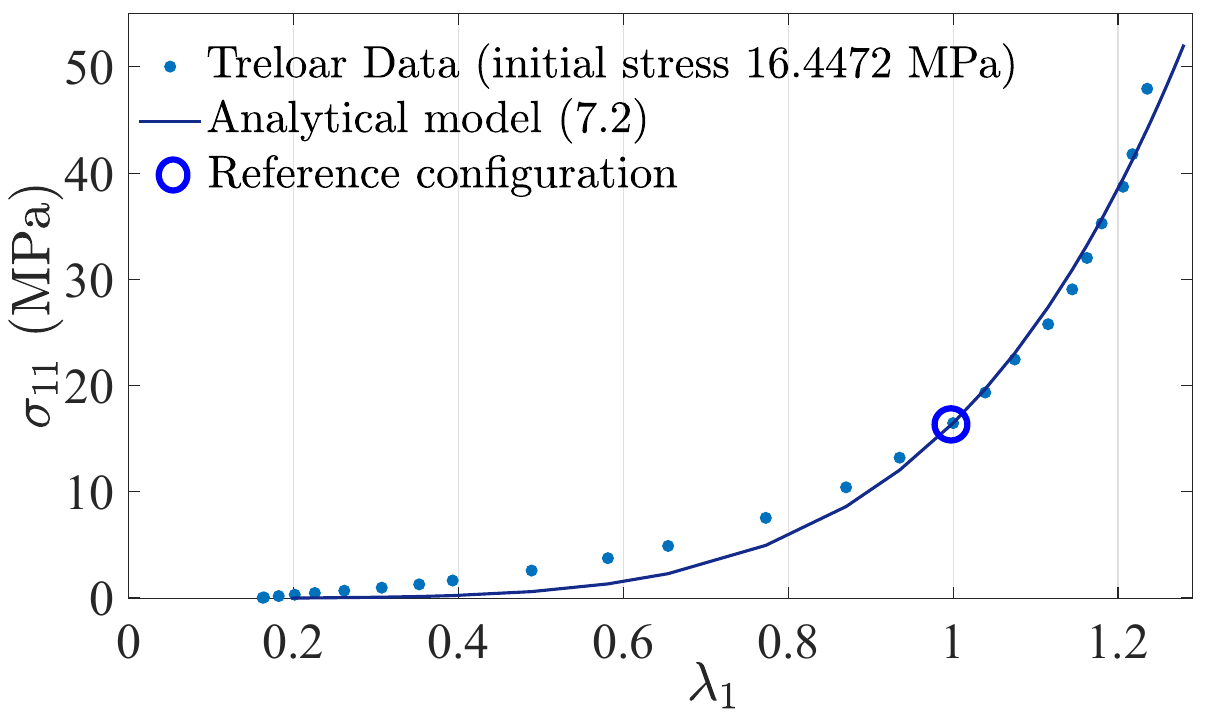}\
    \includegraphics[width=0.48\linewidth]{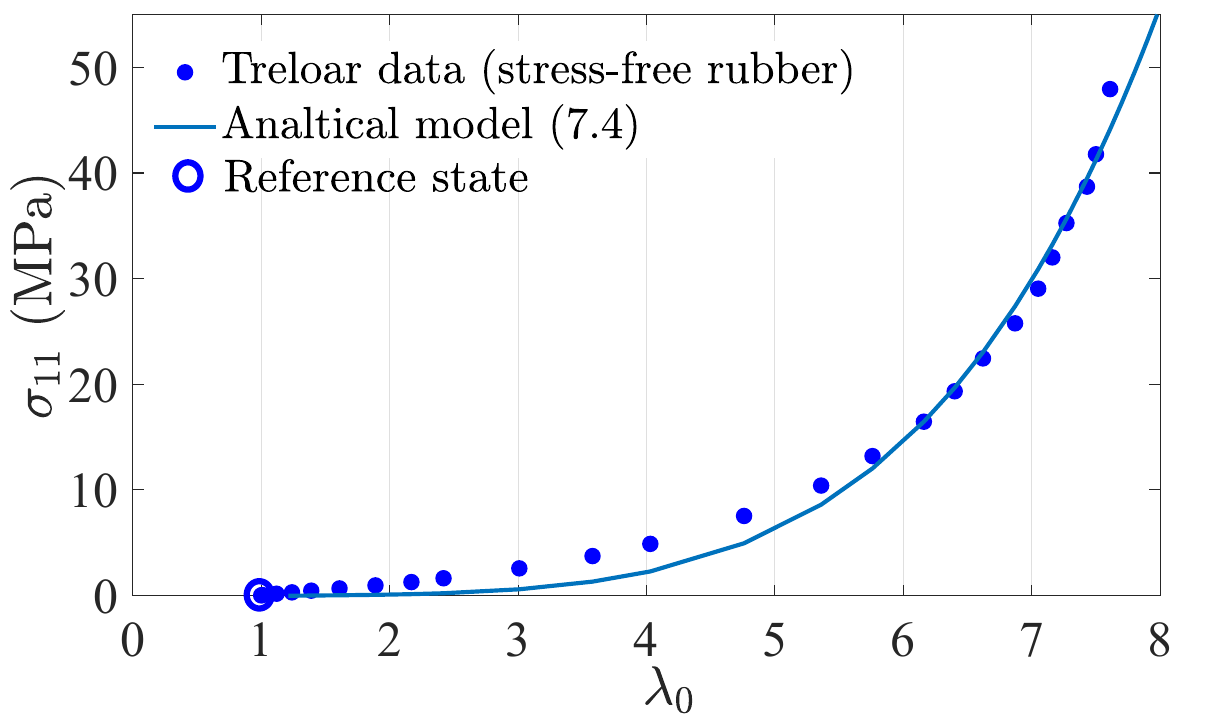}\\
    \includegraphics[width=0.48\linewidth]{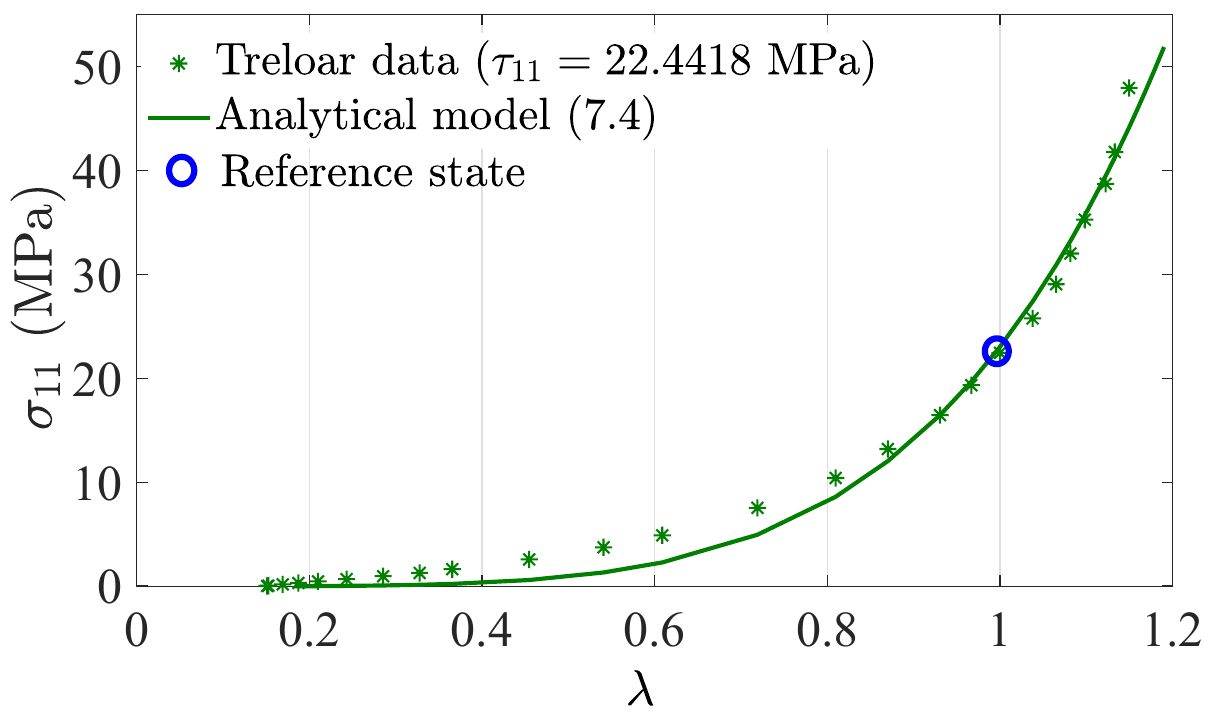}\
    \includegraphics[width=0.48\linewidth]{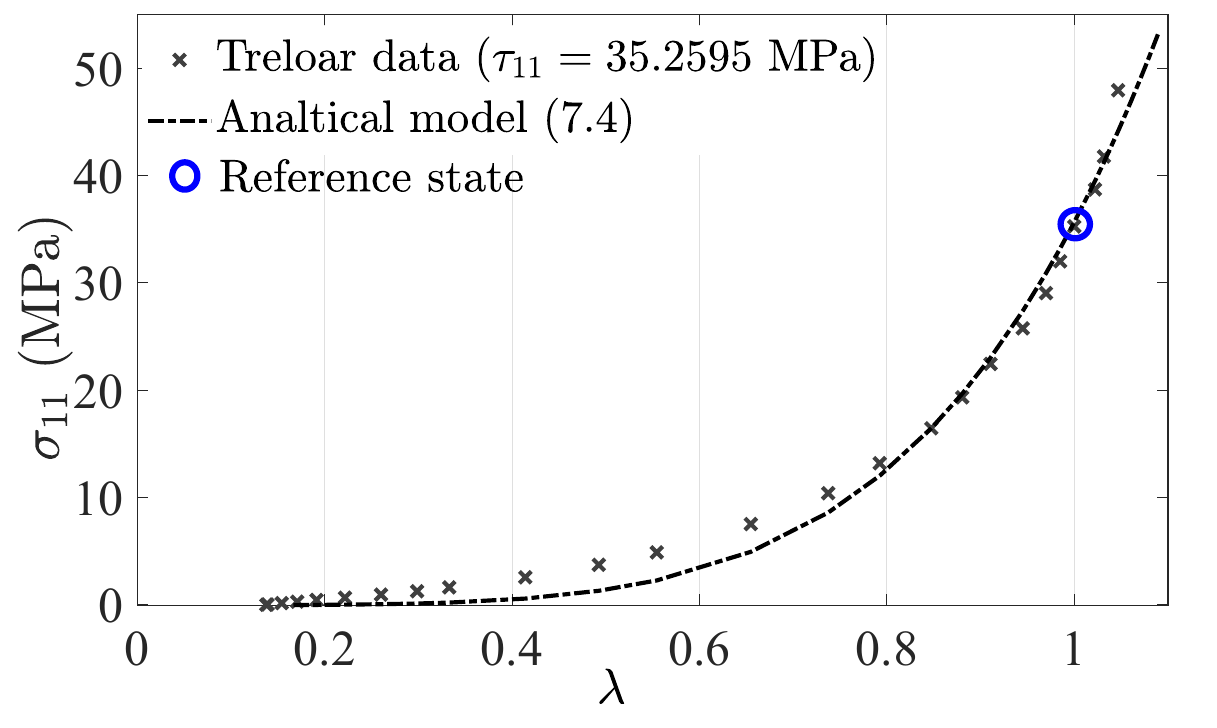}
    \caption{\textcolor{black}{Comparison of Treloar data (Table \ref{tab:my_label}--Table \ref{tab4}) with analytical models for various initially stressed reference configurations. \textbf{Top left:} The constitutive relation \eqref{ana} is validated against the experimental data in Table \ref{tab2} for the reference configuration $\mathfrak{R}_1$ with an initial stress of 16.4472 $\mathrm{MPa}$. The optimized parameters $\mu_1$ and $q_1$, provided in \eqref{expt_par} for the reference $\mathfrak{R}_1$, are then utilized to predict the response of all other references using \eqref{immm} and \eqref{exp}. \textbf{Top right, bottom left, and bottom right:} Comparison of the predicted response \eqref{exp} for the stress-free reference (top right) and references with initial stresses of $\tau_{11} = 22.4418$ $\mathrm{MPa}$ (bottom left) and $\tau_{11} = 35.2595$ $\mathrm{MPa}$ (bottom right). Each figure highlights the reference configurations. All figures exhibit identical agreement with experimental results, despite the optimization being performed exclusively for the top-left case.}}
    \label{fig:enter-label}
\end{figure}

\textcolor{black}{Treloar conducted a well-known experiment using an initially stress-free state of rubber. In Section \ref{Data_Traloar}, we transform the experimental results to various initially stressed references. The response of reference \( \mathfrak{R}_1 \) is validated using the two-parameter model \eqref{vu-1} in Section \ref{corroboration}. The optimized parameters \( \mu_1 \) and \( q_1 \) for reference \( \mathfrak{R}_1 \) are then applied to derive the constitutive relation in Eqn. \eqref{developed}, which is directly compared with the experimental results for all reference configurations \( \mathfrak{R}_0 \) and \( \mathfrak{R} \).
\subsection{Transforming Treloar data to various initially stressed reference configurations}
\label{Data_Traloar}
Table \ref{tab:my_label} presents Treloar experimental data \cite{treloar1944stress} in terms of the principal stretch $\left(\lambda_0\right)$ and the corresponding Cauchy stress (in MPa) for uniaxial tension of initially stress-free rubber. For a uniaxial Cauchy stress of $16.4472$ $\mathrm{MPa}$, the principal stretch measured from the stress-free state was recorded as
\begin{equation}
\hat{\lambda}=6.16.\label{lamh}
\end{equation}
At this point, we choose this state as the initially stressed reference $\mathfrak{R}_1$, where the axial component of the initial stress $\bftau_1$ is set to $16.4472$ $\mathrm{MPa}$. The principal stretch from this reference $\mathfrak{R}_1$ is obtained by scaling the principal stretch values in Table \ref{tab:my_label} using the relation $\lambda_1=\tfrac{\lambda_0}{\hat{\lambda}}$, where $\hat{\lambda}$ is defined in \eqref{lamh}. The resulting stretch-stress data for reference $\mathfrak{R}_1$ are presented in Table \ref{tab2}. This dataset is used to optimize the material parameters in the model \eqref{vu-1} in Section \ref{corroboration}.
\begin{table}[hbt!]
\parbox{.45\linewidth}{
    \centering
    \begin{tabular}{|c|c|}
    \hline
    Principal stretch $\left(\lambda_0\right)$ & Cauchy stress (MPa)\\
    \hline
\textbf{1.000000}&                    \textbf{0.0000}\\
\hline
 1.010000&     0.0303\\
 \hline
     1.120000&     0.1568\\
     \hline
     1.240000 &     0.2852\\
     \hline
     1.390000 &     0.4448\\
     \hline
     1.610000&     0.6601\\
     \hline
     1.890000&     0.9450\\
     \hline
     2.170000&     1.2586\\
     \hline
     2.420000&     1.6214\\
     \hline
     3.010000&     2.5585\\
     \hline
     3.580000&     3.7232\\
     \hline
     4.030000&     4.8763\\
     \hline
     4.760000&     7.5208\\
     \hline
     5.360000&     10.3984\\
     \hline
     5.760000&     13.1904\\
     \hline
     6.160000&     16.4472\\
     \hline
     6.400000&     19.3280\\
     \hline
     6.620000&     22.4418\\
     \hline
     6.870000&     25.7625\\
     \hline
     7.050000&     29.0460\\
     \hline
     7.160000&     32.0052\\
     \hline
     7.270000&     35.2595\\
     \hline
     7.430000&     38.7103\\
     \hline
     7.500000&     41.7750\\
     \hline
     7.610000&     47.9430\\
     \hline
    \end{tabular}
    \caption{Uni-axial tension Treloaor data for stress-free rubbers}
    \label{tab:my_label}}\quad
    \parbox{.45\linewidth}{
    \centering
    \begin{tabular}{|c|c|}
    \hline
    Principal stretch $\left(\lambda_1\right)$ & Cauchy stress (MPa)\\
    \hline
0.162338&                    0.0000\\
\hline
 0.163961&     0.0303\\
 \hline
    0.181818&     0.1568\\
     \hline
     0.201299 &     0.2852\\
     \hline
     0.225649 &     0.4448\\
     \hline
     0.261364&     0.6601\\
     \hline
     0.306818&     0.9450\\
     \hline
     0.352273&     1.2586\\
     \hline
     0.392857&     1.6214\\
     \hline
     0.488636&     2.5585\\
     \hline
     0.581168&     3.7232\\
     \hline
     0.654220&     4.8763\\
     \hline
     0.772727&     7.5208\\
     \hline
     0.870129&     10.3984\\
     \hline
     0.935064&     13.1904\\
     \hline
     \textbf{1.000000}&     \textbf{16.4472}\\
     \hline
     1.038961&     19.3280\\
     \hline
     1.074675&     22.4418\\
     \hline
     1.115259&     25.7625\\
     \hline
     1.144480&     29.0460\\
     \hline
     1.162337&     32.0052\\
     \hline
     1.180194&     35.2595\\
     \hline
     1.206168&     38.7103\\
     \hline
     1.217532&     41.7750\\
     \hline
    1.235389&     47.9430\\
     \hline
    \end{tabular}
    \caption{Uniaxial tension data for the reference $\mathfrak{R}_1$ with uniaxial initial tension of $16.4472\mathrm{MPa}$. The reference state is emphasized with bold-fonts.}
    \label{tab2}}
\end{table}
We also present the Treloar data for other initially stressed references in Tables \ref{tab3}-\ref{tab4}.
\subsection{Determining material parameters for reference $\mathfrak{R}_1$ and comparing the developed model for references $\mathfrak{R}$}
\label{corroboration}
In this section, we first characterize the experimental results (Table \ref{tab2}) for the initially stressed reference $\mathfrak{R}_1$ using the model presented in Eqn. \eqref{vu-1}. The principal Cauchy stress component for uniaxial tension, as derived from Eqn. \eqref{vu-1},  is given by,
\begin{equation} \sigma_{11}=(\mu_1+16.4472)\lambda_1^{q_1}-\mu_1\frac{1}{\lambda_1^{q_1/2}},\label{ana}
\end{equation}
where $16.4472$ $\mathrm{MPa}$ is the magnitude of uniaxial initial stress in $\mathfrak{R}_1$. An optimization function in \emph{MATLAB} is employed to determine the following optimum material parameters:
\begin{align}
   & \mu_1=1\times 10^{-3}\, \mathrm{MPa} && q_1=4.6702.\label{expt_par}
\end{align}
Figure \ref{fig:enter-label} (top left) presents a comparison between the experimental data from Table \ref{tab2} and the analytical model in Eqn. \eqref{ana}, using the optimized parameters from Eqn. \eqref{expt_par}.}
\textcolor{black}{The reference state is highlighted and encircled in the figure. A satisfactory agreement is observed between the experimental data and the optimized analytical model, as the constitutive relation \eqref{vu-1} corresponds to the single-term Ogden model for the stress-free reference (Section \ref{stressfree_example}). This agreement can certainly be improved by slightly increasing the number of parameters in the model. More details of single-term and multiple-term Ogden models can be found in \cite{steinmann2012hyperelastic}.}

\textcolor{black}{Next, we determine the response of all other references: $\mathfrak{R}$ and $\mathfrak{R}_0$.
To achieve this, the optimal material parameters $\mu_1$ and $q_1$ for $\mathfrak{R}_1$ \eqref{expt_par} are substituted into Eqn. \eqref{immm} to compute $\textcolor{black}{p_1}$. This parameter is then used to directly evaluate the Cauchy stress (Eqn. \eqref{developed}) for the reference configurations $\mathfrak{R}_0$ and $\mathfrak{R}$. For uniaxial tension, the principal Cauchy stress is given by Eqn. \eqref{developed} as
\begin{equation} \sigma_{11}=\left(\tau_{11}+\textcolor{black}{p_1}\right)\lambda^{q_1}-\textcolor{black}{p_1}\frac{1}{\lambda^{q_1/2}}.\label{exp} \end{equation}}
\begin{figure}[hbt!]
    \centering
    \includegraphics[width=\linewidth]{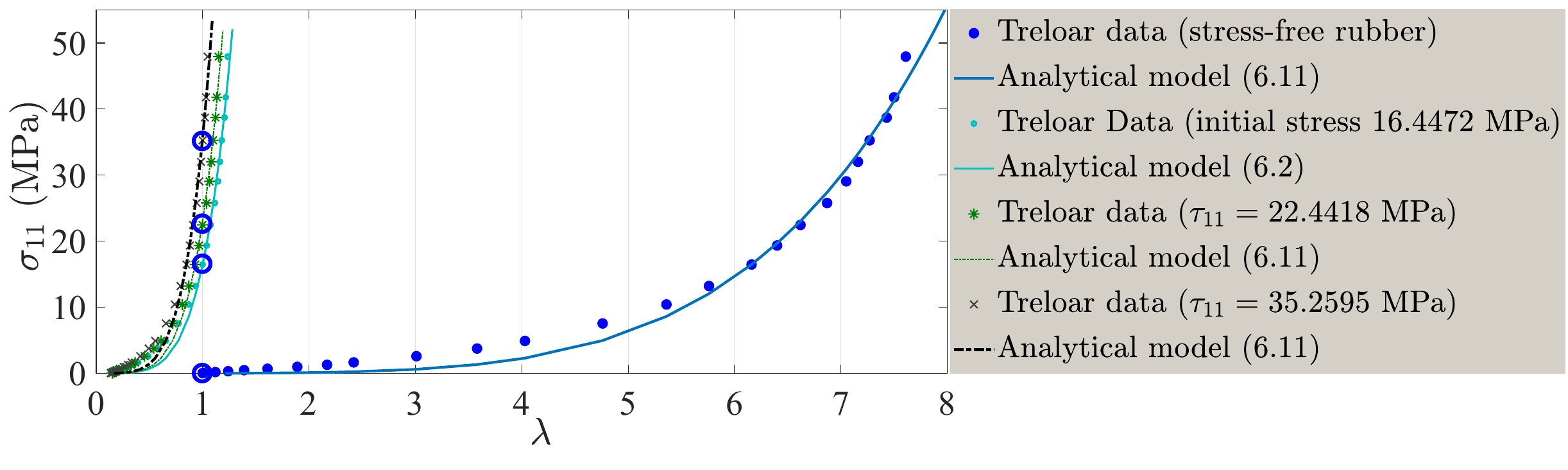}
    \caption{\textcolor{black}{Comparing the stress-strain response under varying initial stress levels.}}
    \label{fig:1-label}
\end{figure}
\textcolor{black}{Figure \ref{fig:enter-label} compares the analytical prediction from Eqn. \eqref{exp} with experimental data from different reference configurations $\mathfrak{R}_0$ and $\mathfrak{R}$. The experimental results for a stress-free reference (Table \ref{tab:my_label}), for $\tau_{11}=22.4418$ $\mathrm{MPa}$ (Table \ref{tab3}), and for $\tau_{11}=35.2595$ $\mathrm{MPa}$ (Table \ref{tab4}), are compared with analytical predictions in Figure \ref{fig:enter-label} (top right), (bottom left), and (bottom right) respectively.
In each figure, the reference configuration is highlighted and encircled for clarity.}

\textcolor{black}{Figure \ref{fig:enter-label} illustrates that the comparisons of the developed analytical model with the experimental results for all reference configurations are consistent, identical, and equivalent. This observation validates that the proposed framework for altering the reference preserves the inherent aspects and fundamental characteristics of the response.}

\textcolor{black}{We note here that the responses illustrated in Figure \ref{fig:enter-label} are presented with varying horizontal axis ranges. To examine the influence of initial stress more comprehensively, all responses are plotted together in Figure \ref{fig:1-label} using a common set of coordinates.}
\section{The dependency and independence in the functional form of the constitutive relation on the reference}
\label{ogd-art}
There has been some debate on whether the functional form of the constitutive relation depends on the choice of reference \cite{10.1093/qjmam/hbx020,ogden2023change}. Addressing this question is not the focus of this paper; our main objective is to develop methodologies for transforming the reference configuration. However, in this section, we briefly demonstrate that the proposed constitutive framework is consistent with all existing theories on this matter.

So far, we have not described the Cauchy stress $\bfsigma$ as a function $\bfsigma\left(.,.\right)$ or $\bfsigma_1\left(.,.\right)$. We now describe the constitutive relation Eqn. \eqref{actual} by the function
\begin{equation}
\bfsigma_1\left(\bftau_1,\bfF_0\right)=\Psi_1\bfF_0\bfF_0^T+\bfF_0\bftau_1\bfF_0^T-p\bfI,\label{rtr}
\end{equation}
where $\bfF_0=\bfF{{\bfF}}_1$. Similarly, we represent the relation Eqn. \eqref{actual---111} using the function
\begin{equation}
\bfsigma\left(\bftau,\bfF\right)=p_1\bfF\bfF^T+\bfF\tau\bfF^T-p\bfI.\label{rtr1}
\end{equation}
Each of the parameters in the above expressions may exhibit spatial variation. Both \eqref{actual} and \eqref{actual---111} denote the Cauchy stress in the configuration $\mathfrak{C}$, Thus, it follows that
\begin{equation}
\bfsigma_1\left(\bftau_1,\bfF_0\right)=\bfsigma\left(\bftau,\bfF\right),\quad \text{where }\bfF_0=\bfF\boldsymbol{\tilde{F}}.\label{ogd}
\end{equation}
On the other hand, from Eqn. \eqref{git}, we have $p_1=\Psi_1$ for $\bftau=\bftau_1$. Substituting the above into Eqn. \eqref{rtr1}, we obtain
\begin{equation}
\bfsigma\left(\bftau_1,\bfF_0\right)=\Psi_1\bfF_0\bfF_0^T+\bfF_0\bftau_1\bfF_0^T-p\bfI,\label{rtr--1}
\end{equation}
Comparing Eqn. \eqref{rtr} with Eqn. \eqref{rtr--1}, we obtain
\begin{equation}
\bfsigma_1\left(\bftau_1,\bfF_0\right)=\bfsigma\left(\bftau_1,\bfF_0\right).\label{rrtr}
\end{equation}
Substituting Eqn. \eqref{rrtr} into Eqn. \eqref{ogd}, we obtain
\begin{equation}
\bfsigma\left(\bftau_1,\bfF_0\right)=\bfsigma\left(\bftau,\bfF\right).\label{isri}
\end{equation}
Therefore, the present case agrees with both the representations in Eqn. \eqref{isri} \cite{10.1093/qjmam/hbx020} and Eqn. \eqref{ogd} \cite{ogden2023change}. Eqn. \eqref{ogd} has been recently introduced. However, the method to determine the constitutive relation $\bfsigma\left(\bfF,\bftau\right)$ that directly satisfies \eqref{ogd} \cite{ogden2023change} was not known. The present work attempts to resolve this question.
It is interesting to note here that the existing \cite{johnson1995use,saravanan2008representation,gower2015initial,mukherjee2022representing} approaches to changing the reference configuration will remain unchanged whether Eqn. \eqref{ogd}, or  Eqn. \eqref{isri} is considered. 


We further note that Eqn. \eqref{isri} can be rewritten as
\begin{equation} \bfsigma\left(\bftau_1,\bfF{\bfF}_1\right)=\bfsigma\left(\bftau,\bfF\right), \label{symm?} \end{equation}where the left-hand side of \eqref{symm?} represents the constitutive relation relative to the reference $\mathfrak{R}_1$, while the right-hand side corresponds to the constitutive relation for the reference $\mathfrak{R}$.  As Rajagopal and Wineman \cite{rajagopal2024residual} have shown that ``\emph{one can only discuss the material symmetry of the body with regard to a specific configuration}", Eqn. \eqref{symm?} should not be applied for determining the symmetry of either of the two references $\mathfrak{R}_1$, or $\mathfrak{R}$.
\section{Conclusions}
The response of the stress-free material is commonly used to determine the
constitutive relations for an initially stressed reference configuration. This
process of changing the reference configuration always starts from a
stress-free
configuration~\cite{johnson1995use,gower2015initial,saravanan2008representation,mukherjee2022representing,mukherjee2022influence}.
This paper initiates the process of changing the reference configuration from a specified initial stressed state. We
employ the known response of a given initially stressed reference to determine
the response of the stress-free material and that of any other arbitrary
initially stressed references. Two general approaches and one special approach
are presented. {A strain energy, based on generalized
invariants, is developed and successfully corroborated using experimental
results.} An important part of the paper is the formulation of universal
relations to general implicit elasticity. 

The first approach directly transforms a specific constitutive relation from one stressed reference configuration to another.

The second approach is broad and general. In this approach, the constitutive relations for the stress-free reference are determined first. We then use the response of the stress-free material to uncover the response of any initially stressed reference. It is observed that altering the reference configuration may lead to an implicit constitutive relation for other reference configurations, while the response of the known initially stressed reference is governed by Green elasticity. This establishes a direct correlation between implicit elasticity and Green elasticity.

Three explicit constitutive relations have also been established for various initially stressed and the stress-free references. {For one of these relations, the free energy relative to the known reference configuration is expressed in terms of invariants involving initial stress and the fractional powers of the right Cauchy stretch tensor. It is found out that the response of the associated stress-free reference is characterized by the Ogden model.}

The third approach extends the first approach to create a general model applicable to any arbitrary initially stressed reference. The initial stress in the known reference is systematically transformed to introduce an \emph{imaginary} configuration. This facilitates the inversion of a new family of constitutive relations.

{We use the constitutive relation involving generalized invariants to validate the experimental results. To prepare the necessary experimental data, we transform the Treloar data for uniaxial tension to various initially stressed references. For a particular stressed reference, denoted as $\mathfrak{R}_1$, the prepared experimental result is corroborated to optimize a two-parameter model. We use the optimal parameters for the reference $\mathfrak{R}_1$ to obtain the constitutive relations for all other references, which are directly compared with the respective experimental results. It is noteworthy that, despite the model being optimized solely for the reference configuration $\mathfrak{R}_1$, it provides an identical fit for all references. This demonstrates that our methods for changing the reference configuration do not alter the essential features of the constitutive relation. As a result, a consistently good agreement with all the experimental data for various initially stressed reference configurations was achieved by optimizing only two parameters.}

This paper explores and establishes the relationship between implicit and Green elasticity. It examines the responses of various initially stressed and stress-free references, develops new approaches for changing reference configurations, {and validates these approaches using experimental results.} Furthermore, the paper establishes universal relations within general implicit elasticity, which can facilitate the application of general implicit constitutive relations to a wide range of problems.
\section*{Acknowledgements}
The author expresses sincere gratitude to Prof.~Parag Ravindran (Indian Institute of Technology Madras) and Prof.~C. S. Jog (Indian Institute of Science Bangalore) for critical remarks. The author is grateful to the anonymous reviewer for recommending an experimental validation, which enhances the manuscript.

The author acknoledges the financial support provided by Indian Institute of Technology Palakkad, via Faculty Seed Grant.
\appendix
\section{Universal relations in implicit elasticity}
\label{uni}
\textcolor{black}{In this section, we try to establish the universal relation for implicit elasticity, as follows. Eqn. \eqref{consti-7} is rewritten below for completeness: 
\begin{align}
    \bf0& = \bar{\Psi}_1 \bfI + \bar{\Psi}_2 \hat{\bfB}^{-1} +\bar{\Psi}_7 \bftau_1+\bar{\Psi}_8\left(\bftau_1\hat{\bfB}^{-1}+\hat{\bfB}^{-1}\bftau_1\right)+\bar{\Psi}_9 \bftau_1^2+\bar{\Psi}_{10}\left(\bftau_1^2\hat{\bfB}^{-1}+\hat{\bfB}^{-1}\bftau_1^2\right)-\bar{p}\hat{\bfB}.\label{consti-71}
\end{align}
Post-multiplying both sides of \eqref{consti-71} with $\hat{\bfB}^n$, we obtain
\begin{align}
    \bf0& = \bar{\Psi}_1 \hat{\bfB}^n + \bar{\Psi}_2 \hat{\bfB}^{n-1} +\bar{\Psi}_7 \bftau_1\hat{\bfB}^n+\bar{\Psi}_8\left(\bftau_1\hat{\bfB}^{n-1}+\hat{\bfB}^{-1}\bftau_1\hat{\bfB}^{n}\right)+\bar{\Psi}_9 \bftau_1^2\hat{\bfB}^n+\bar{\Psi}_{10}\left(\bftau_1^2\hat{\bfB}^{n-1}+\hat{\bfB}^{-1}\bftau_1^2\hat{\bfB}^n\right)-\bar{p}\hat{\bfB}^{n+1}.\label{consti-}
\end{align}
The anti-symmetric part of Eqn. \eqref{consti-} provides
\begin{align}
    \bf0& = \bar{\Psi}_7\left( \bftau_1\hat{\bfB}^n-\hat{\bfB}^n\bftau_1\right)+\bar{\Psi}_8\left( \bftau_1\hat{\bfB}^{n-1}-\hat{\bfB}^{n-1}\bftau_1\right)+\bar{\Psi}_8\left(\hat{\bfB}^{-1}\bftau_1\hat{\bfB}^n-\hat{\bfB}^n\bftau_1\hat{\bfB}^{-1}\right)+\bar{\Psi}_9\left( \bftau_1^2\hat{\bfB}^n-\hat{\bfB}^n\bftau_1^2\right)\nonumber\\&+\bar{\Psi}_{10}\left( \bftau_1^2\hat{\bfB}^{n-1}-\hat{\bfB}^{n-1}\bftau_1^2\right)+\bar{\Psi}_{10}\left(\hat{\bfB}^{-1}\bftau_1^2\hat{\bfB}^n-\hat{\bfB}^n\bftau_1^2\hat{\bfB}^{-1}\right).\label{consti---}
\end{align}
Since \eqref{consti---} should be satisfied for all $n$, it appears reasonable to consider that universal relations:
\begin{align}
&\bftau_1\hat{\bfB}^n=\hat{\bfB}^n\bftau_1 && \bftau_1^2\hat{\bfB}^n=\hat{\bfB}^n\bftau_1^2
\end{align}
hold for implicit elasticity.}
\section{Influence of non-unique rotation on the constitutive relation for the stress-free reference}
\label{apex}
A closer look into Figure \ref{fig1} will show that any additional arbitrary rotation $\bfQ$ keeps the stress-free configuration $\mathfrak{R}_0$ stress-free. Considering this additional rotation, Equation \eqref{ff} can be expressed in a more general form as 
 \beq\bfQ\bfF_1=\hat{\bfF}^{-1}.\eeq This non-unique rotation $\bfQ$ changes Equation (\ref{bb}b) to \beq\bfQ\bfB_1\bfQ^T=\hat{\bfC}^{-1}.\eeq However, this rotation does not alter any other equation in Section \ref{1st}.
\textcolor{black}{\section{Treloar data from various stressed reference configurations}
This section presents the Treloar uniaxial test transformed into two initially stressed references in continuation with Section \ref{Data_Traloar}.
\begin{table}[hbt!]
\parbox{.45\linewidth}{
    \centering
    \begin{tabular}{|c|c|}
    \hline
    Principal stretch $\left(\lambda\right)$ & Cauchy stress (MPa)\\
    \hline
   0.151057&                   0.0000\\
   \hline
   0.152568&   0.0303\\
   \hline
   0.169184&   0.1568\\
   \hline
   0.187311&   0.2852\\
   \hline
   0.209970&   0.4448\\
   \hline
   0.243202&   0.6601\\
   \hline
   0.285498&   0.9450\\
   \hline
   0.327794&   1.2586\\
   \hline
   0.365559&   1.6214\\
   \hline
   0.454683&   2.5585\\
   \hline
   0.540785&   3.7232\\
   \hline
   0.608761&   4.8763\\
   \hline
   0.719033&   7.5208\\
   \hline
   0.809668&  10.3984\\
   \hline
   0.870090&  13.1904\\
   \hline
   0.930514&  16.4472\\
   \hline
   0.966767&  19.3280\\
   \hline
   \textbf{1.000000}&  \textbf{22.4418}\\
   \hline
   1.037764&  25.7625\\
   \hline
   1.064955&  29.0460\\
   \hline
   1.081571&  32.0052\\
   \hline
   1.098187&  35.2595\\
   \hline
   1.122356&  38.7103\\
   \hline
   1.132931&  41.7750\\
   \hline
   1.149547&  47.9430\\
     \hline
    \end{tabular}
    \caption{Uniaxial tension data for the reference $\mathfrak{R}$ with uniaxial initial tension of $22.4418\mathrm{MPa}$. The reference state is emphasized with bold-fonts.}
    \label{tab3}}\quad
\parbox{.45\linewidth}{
    \centering
    \begin{tabular}{|c|c|}
    \hline
    Principal stretch $\left(\lambda\right)$ & Cauchy stress (MPa)\\
\hline
   0.137551&             0.0000\\
   \hline
   0.138927&   0.0303\\
   \hline
   0.154057&   0.1568\\
   \hline
   0.170563&   0.2852\\
   \hline
   0.191196&   0.4448\\
   \hline
   0.221458&   0.6601\\
   \hline
   0.259972&   0.9450\\
   \hline
   0.298486&   1.2586\\
   \hline
   0.332874&   1.6214\\
   \hline
   0.414030&   2.5585\\
   \hline
   0.492435&   3.7232\\
   \hline
   0.554333&   4.8763\\
   \hline
   0.654746&   7.5208\\
   \hline
   0.737276&  10.3984\\
   \hline
   0.792297&  13.1904\\
   \hline
   0.847318&  16.4472\\
   \hline
   0.880330&  19.3280\\
   \hline
   0.910591&  22.4418\\
   \hline
   0.944979&  25.7625\\
   \hline
   0.969739&  29.0460\\
   \hline
   0.984869&  32.0052\\
   \hline
   \textbf{1.000000}&  \textbf{35.2595}\\
   \hline
   1.022008&  38.7103\\
   \hline
   1.031637&  41.7750\\
   \hline
   1.046768&  47.9430\\
     \hline
    \end{tabular}
    \caption{Uniaxial tension data for the reference $\mathfrak{R}$ with uniaxial initial tension of $35.2595\mathrm{MPa}$. The reference state is emphasized with bold-fonts.}
    \label{tab4}}
\end{table}}
\newpage
\bibliographystyle{unsrt}
\bibliography{main}
\end{document}